\documentclass[iop,appendixfloats]{emulateapj}

\usepackage{amssymb,gensymb,graphics,textcomp}
\usepackage[section] {placeins}

\newcommand{\OI}{[\ion{O}{1}]}

\newcommand{\OIII}{[\ion{O}{3}]}
\newcommand{\SII}{[\ion{S}{2}]}
\newcommand{\NII}{[\ion{N}{2}]}

\newcommand{\OII}{[\ion{O}{2}]}

\newcommand{\Ha}{H$\alpha\,$}
\newcommand{\Hb}{H$\beta\,$}

\shorttitle{}
\shortauthors{Rich et al.}

\begin{document}

\title{An Integral Field Study of Abundance Gradients in Nearby LIRGs}

\author{J. A. Rich \altaffilmark{1}, P. Torrey\altaffilmark{2}, L. J. Kewley\altaffilmark{1}, M. A. Dopita\altaffilmark{1,}\altaffilmark{3,}\altaffilmark{4}, \& D. S. N. Rupke\altaffilmark{5}  }
\email{jrich@ifa.hawaii.edu}
\altaffiltext{1}{Institute for Astronomy, University of Hawaii, 2680 Woodlawn Drive, Honolulu, HI 96822}
\altaffiltext{2}{Harvard Smithsonian Center for Astrophysics, 60 Garden St., Cambridge, MA 02138}
\altaffiltext{3}{Research School of Astronomy and Astrophysics, Australian National University, Cotter Rd., Weston ACT 2611, Australia}
\altaffiltext{4}{Astronomy Department, Faculty of Science, King Abdulaziz University, PO Box 80203, Jeddah, Saudi Arabia}
\altaffiltext{5}{Department of Physics, Rhodes College, Memphis, TN 38112}

\date{\today}

\begin{abstract}
We present for the first time metallicity maps generated using data from the Wide Field Spectrograph (WiFeS) on the ANU 2.3m of 9 Luminous Infrared Galaxies (LIRGs) and discuss the abundance gradients and distribution of metals in these systems. We have carried out optical integral field spectroscopy (IFS) of several several LIRGs in various merger phases to investigate the merger process.  In a major merger of two spiral galaxies with preexisting disk abundance gradients, the changing distribution of metals can be used as a tracer of gas flows in the merging system as low metallicity gas is transported from the outskirts of each galaxy to their nuclei. We employ this fact to probe merger properties by using the emission lines in our IFS data to calculate the gas-phase metallicity in each system. We create abundance maps and subsequently derive a metallicity gradient from each map. We compare our measured gradients to merger stage as well as several possible tracers of merger progress and observed nuclear abundances. We discuss our work in the context of previous abundance gradient observations and compare our results to new galaxy merger models which trace metallicity gradient. Our results agree with the observed flattening of metallicity gradients as a merger progresses. We compare our results with new theoretical predictions that include chemical enrichment. Our data show remarkable agreement with these simulations.\\
\end{abstract}

\keywords{Galaxies: Abundances, Galaxies: Interactions, Infrared: Galaxies}

\section{Introduction}
Galaxy collisions and mergers represent a key stage in the evolution of galaxies in the local universe and beyond. Throughout cosmic time gravitational forces have assembled ever larger galactic systems from the collisions and mergers of smaller fragments. Beginning with the theoretical \citet{Toomre77} sequence of merging galaxies, theoretical modeling of massive, merging galaxies now includes detailed physics, allowing us to make predictions about galaxy evolution over cosmic time.  The history of chemical enrichment is tied to both star formation and the dynamic redistribution of gas throughout the lifetime of a galaxy and is drastically modified by galaxy merger events \citep{Kobayashi04,Rupke10a,Torrey11}.

Theory predicts that major mergers encourage the formation of bars in the stellar and gas disks, which induce vigorous gas inflows as the gas looses angular momentum to the stellar component \citep{Barnes96}.  These inflows are thought to be responsible for fueling a massive central starburst and feeding AGN and/or quasar activity \citep{Mihos96,Barnes96}. For a spiral galaxy with a preexisting metallicity gradient gas inflow flattens the gradient by diluting the higher abundance gas in the central regions with the lower abundance gas from the outer parts of the galaxy \citep{Rupke10b,Rupke10a,Kewley10}. This flattening is compounded as the spiral arms are stretched by tidal effects \citep{Torrey11}. Here we investigate this process in nearby Luminous Infrared Galaxies (LIRGs)

LIRGs are excellent targets to inform the study of galaxy evolution. Although relatively rare, LIRGs compose the larger part of the IR luminosity of the universe by z$\sim$1 \citep{lefloch05}. As a group LIRGs span merger stages from isolated to post-merger and contain massive starbursts, AGN, shocks and a variety of stellar population ages \citep{Sanders96,Kewley01b,Sanders03,Yuan10}. In most cases the starbursts and AGN in LIRGs are driven by ongoing mergers and the progression of these mergers is correlated with a rise in the IR luminosity, with massive merging LIRGs eventually surpassing $L_{IR}=10^{12} L_{\sun}$ to become Ultraluminous Infrared Galaxies (ULIRGs) (e.g. \citealt{Sanders88}). Additionally, both merger activity and the incidence of U/LIRGs increase at higher redshift ~\citep{deRavel09, Bundy09}. It is clear that merging, IR-luminous systems played an important role during the peak of star formation and in the assembly of present-day massive galaxies. This makes LIRGs an ideal target to study the mixing and re-distribution of heavy elements in mergers to test the predictions of merger models.

~\citet{Kewley06b} found the first observational link between merging spiral galaxies and tidally induced gas flow in the depressed nuclear metallicities of close merging pairs. Other studies of mass-metallicity and luminosity-metallicity correlations find that merging systems tend to be underabundant for their size and brightness \citep{Lee04,Rupke08,Ellison08}.  Theoretical modeling of the metal distribution in merging systems reproduces the observed nuclear underabundance and predicts a flattening of a preexisting abundance gradient as the merger progresses  ~\citep{Rupke10a}. Recent multi-slit spectroscopy of HII regions in close-pair spiral galaxies confirmed that gas-phase metallicity gradients are indeed flatter in merging systems when compared to a control sample of isolated spirals ~\citep{Kewley10,Rupke10b}.

In this paper we present an integral field spectroscopic (IFS) study of chemical abundances in 9 nearby LIRGs. Our systems are a subset of a larger IFS sample of nearby U/LIRGs: the Wide Field Spectrograph (WiFeS) Integral Field Unit (IFU) Great Observatory All-Sky LIRG Survey (GOALS) Sample (WIGS). IFS data provides a wealth of useful information about nearby galaxies, allowing the observer to measure several aspects of the gas physics and stars over large areas. By using our IFS data to generate maps of emission line fluxes, we can similarly create abundance maps and track metallicity gradient as a function of merger stage and sample later merger stages than previously considered. 
 
We provide a summary of our sample, the observations and the data reduction in Section 2. In Section 3 we discuss the analysis of our spectra and our metallicity determinations and calibrations. We also present the resulting metallicity maps and gradients created from our IFU data in Section 3. We analyze our observations and compare our measurements with previous observations of metallicities in merging systems in Section 4. In Section 5 we discuss our observations in the context of recent merger models. Section 6 gives our conclusions. Throughout this paper we adopt the cosmological parameters H$_{0}$=70.5~km~s$^{-1}$Mpc$^{-1}$, $\Omega_{\mathrm{V}}$=0.73, and $\Omega_{\mathrm{M}}$=0.27, based on the five-year WMAP results \citet{Hinshaw09} and consistent with the \citet{Armus09} summary of the GOALS sample.

\section{Sample, Observations and Data Reduction}
\subsection{Our Sample}
Our targets are drawn from the Great Observatory All-Sky LIRG Survey (GOALS) sample \citep{Armus09}.  GOALS is a multi-wavelength survey of the brightest 60$\mu$m extragalactic sources in the local universe ($log(L_{IR}/L_{\sun}) > 11.0$) with redshifts z $<$ 0.088 and is a complete subset of the IRAS Revised Bright Galaxy Sample (RBGS) \citep{Sanders03}. Objects in GOALS cover the full range of nuclear spectral types and interaction stages and are excellent analogs for comparison with high-redshift galaxies. Due to the location of WiFeS, our targets comprise a southern sample, with a declination limit of about +15\textdegree.

The LIRGs in our sample represent a variety of merger stages, nuclear separations, luminosities and physical processes. The systems presented in this paper are those objects for which we could measure the extended gas-phase metallicities with sufficient spatial resolution, as the optical emission line gas in these systems is too compact and too extinguished to sufficiently measure a metallicity gradient with our ground-based data. In addition, we rule out galaxies whose spectra are overwhelmingly dominated by non HII-region emission. In our sample this is primarily widespread radiative shocks \citep{Rich11}. Systems ruled out consist primarily of the post-merger objects in our sample. For our sample, the data is insufficient for 10 of the 12 post-merger U/LIRGs in our sample: follow-up on larger telescopes with AO and/or \emph{Hubble Space Telescope} would likely allow for a better investigation of post-merger targest.

We analyze a total of 11 galaxies in 9 systems, shown in Table 1. We classify the merger stage of each system using the scheme adopted by \citet{Yuan10}, which is a slight modification of the merger stage classification outlined by \citet{Veilleux02} based on the comparison of observations with the simulations of \citet{Barnes92,Barnes96}. Isolated systems ('iso') show no signs of interaction and have no companion galaxies within 100 kpc in projected distance. Isolated galaxies may have extremely distant companions beyond 100 kpc, but they show no morphological indication of current or past major interaction. Widely separated systems ('a') show signs of an ongoing major merger including tidal tails and bridges, but companions are separated by a minimum projected distance of 10 kpc.  Closely interacting systems ('b') show more advanced tidal structures and are separated by less than 10 kpc, but are not yet coalesced. Finally we group all coalesced mergers together ('cde'), including diffuse, compact and post-merger systems. All coalesced mergers  are characterized by a single nucleus in ground-based optical and NIR images, with an increasingly compact core as the final stages of the merger take place. Our sample has 2 isolated systems, 1 in stage 'a', 4 in stage 'b' and 2 in stage 'cde'. 

\subsection{Comparison Samples}
We compare our work to the previous samples of~\citet{Kewley10,Rupke10b}.~\citet{Kewley10} study a  sample of 5 pairs of local luminous spiral galaxies from the optically selected galaxy pair sample of ~\citet{Barton00}, each with separations of 15 - 25 kpc. ~\citet{Rupke10b} expand on this sample of interacting systems and also draw from the optically-selected~\citet{Arp66} catalog as well the infrared-selected samples of~\citet{Sanders03} and~\citet{Surace04} for a total sample of 22 interacting galaxies in 9 pairs/groups, all classified in stage 'a' according to the classification used in this paper. ~\citet{Rupke10b} also assemble a control sample of local, isolated spiral galaxies from the literature with properties similar to the spirals in their interacting galaxy sample. Our work overlaps with and extends the Rupke et al. sample to later merger stages. 

The comparison samples consist primarily of gas-rich spirals which are accepted as the progenitors of LIRGs \citep{Barnes96,Mihos96,Veilleux02,Iono04,Ishida04,Naab06,Rupke08}. The higher average L$_{IR}$ of our LIRG sample is the result of intense star formation induced by the merger process-three of the interacting systems in the comparison sample are in fact LIRGs themselves \citep{Rupke10b}. 

Our systems are 1 to 2 magnitudes brighter in M$_{K}$, which generally correlates with stellar mass, as older late-type stars trace the majority of stellar mass with spectral emission peaking in the NIR \citep{Aaronson77}. This enhancement is in part due to the merger process, as the total M$_{K}$ of the later-stage, combined systems will reflect the total M$_{K}$ of the progenitors. An additional enhancement in M$_{K}$ also reflects a  contribution from the intense, young starbursts in LIRGs. Both models and observations show the possibility of a significant contribution to the NIR flux (from 30\% to 60\%) from asymptotic and red giant branch stars \citep{Mouhcine02,Maraston06,Walcher11,Melbourne12}. Thus we expect the M$_{K}$ to be enhanced of order 1 to 2 magnitudes at approximately similar stellar mass indicating the masses of the systems in our sample are roughly consistent with the masses of the comparison sample. This means M$_{K}$ may not correlate well with stellar mass in LIRGs, but the enhancement in M$_{K}$ should still trace the merger process in the same way as L$_{IR}$.

\subsection{Observations \& Data Reduction}
\begin{deluxetable*}{llcccccc} 
\tablewidth{0pc}
\tablecolumns{8} 
\tablenum{1}
\tablehead{ \colhead{IRAS \#}  &  \colhead{Other Name}  & \colhead{log(L$_{IR}$/L$_{\odot}$)} & \colhead{M$_{K}$} & \colhead{PA (\textdegree)} & \colhead{Inclination (\textdegree)} & \colhead{d25 (')} & \colhead{Merger Stage}  }
\centering
\startdata
F01053$-$1746 & IC 1623 A/B                & 11.71 & -24.61 & 63.4  & 34.8 & 1.05 & b \\
~~08355$-$4944 & -                                      & 11.62 & -23.70 & ... & ... & 0.30 & cde\\
F10038$-$3338 & ESO 374-IG032         & 11.78 & ... & ... & ... & 0.76 & cde\\ 
F10257$-$4339 & NGC 3256                   & 11.64 & -24.78  & 83.2 & 48.7 & 3.31 & b\\
F13373$+$0105 E/W & Arp 240                & 11.62 & -24.96/-25.13 & 177.9/85.1 & 34.2/62.1 & 1.48/1.48 & a   \\
F17222$-$5953 & ESO 138-G027            & 11.41 & -24.23 & 41.7 & 51.2 & 1.07 & iso \\
F18093$-$5744 N/S & IC 4687/4689             & 11.62 & -24.47/-23.90 & 51.2/141.4 & 52.7/74.0 & 1.15/0.91 & b \\
F18341$-$5732 & IC 4734                         & 11.35 & -24.48 & 102.6 & 57.4 & 1.32 & iso \\
F19115$-$2124 & ESO 593-IG008           & 11.93 & -25.68 & ... & ... & 0.55 & b 
\enddata

\tablecomments{WiFeS GOALS metallicity sample. Names and IR luminosities are taken from the \citet{Armus09} summary of the GOALS sample. M$_{K}$ is derived using 2mass data. PA, inclination and d25 for each system are all taken from the HyperLeda derived values~\citep{Paturel03} when possible, those with suspect/unavailable values are noted with '...'.  The merger morphology scheme values (adapted from \citet{Yuan10} when available), are given in the final column. Some of the sources classified as ``a'' or ``b'' have multiple pointings of individual galaxies.}  

\end{deluxetable*} 

Our data were taken using WiFeS at the Mount Stromlo and Siding Spring Observatory (SSO) 2.3 m telescope.  WiFeS is a new, dual beam, image-slicing IFU commissioned in May 2009 and described in detail by \citet{Dopita07} and \citet{Dopita10}. Our data consists of blue and red spectra with a resolution of R~3000 \& R~7000 and wavelength coverage of $\sim$3500-5800 \AA~ \& $\sim$5500-7000 \AA~respectively. The data were taken over 5 separate observing runs in July, August and September 2009 and in March and May 2010. 

The data were reduced and flux calibrated using the WiFeS pipeline \citep{Dopita10}. The pipeline uses IRAF routines adapted from the Gemini North's Near-Infrared Integral Field Spectrometer (NIFS) data reduction package. A single WiFeS observation consists of 25, 1\arcsec wide, 38\arcsec long slit spectra with contemporaneous sky spectra if the observation was taken in nod and shuffle mode. All of the final combined data cubes in our sample consist of at least 2 observations of the same target and in some cases a mosaic of two or more pointings. Where there were not sufficient observations to remove cosmic rays via median-combination, cosmic ray removal was performed with the ``dcr'' routine \citep{Pych04}. In this paper we provide only a brief summary of the data reduction, the process is described in detail in \citet{Rich10,Rich11}.

Individual observations are bias-subtracted using bias frames taken as near in time as possible to the observation frames to avoid temporal effects. The observations were flat-fielded using quarts lamp flats and twilight sky flats. The individual spectra are spatially calibrated using a thin wire combined with a continuum lamp. CuAr and NeAr arc lamp spectra taken throughout each night are used to wavelength calibrate each observation. The 25 resulting slitlet-spectra are then rectified into blue-arm and red-arm data cubes and sampled to a common wavelength scale.

Telluric lines are then removed using observations of B-stars or featureless white dwarfs taken at similar airmass the same night as the data. Each data cube is then flux calibrated using observations of flux standard stars taken on the same night. If there were any non-photometric pointings of a target, these pointings were scaled and flux calibrated using photometric data of the same target. We correct for the effects of atmospheric dispersion using the WiFeS pipeline.

Each individual reduced data cube from observations of a single galaxy is then combined to form a final data cube sampled to a common spatial grid for analysis purposes. The WiFeS detectors have 0.5\arcsec pixels along the slit, so data taken in 2010 are binned on-chip by 2 pixels in the spatial direction in order to increase signal to noise and produce square spatial elements $1\arcsec \times 1$\arcsec. Observations taken in 2009 are binned post-reduction to achieve similar results. Typical seeing achieved at SSO during our observations is $\sim1.5$\arcsec, with some variation, on par with the spaxel size for our data cubes. We do not perform further binning (e.g. Voronoi tessellation) prior to our data analysis.

Combined data cubes were aligned astrometrically by comparing a pseudo 'r-band' image generated using the red spectrum from each spectral pixel (spaxel) with either DSS R-band or HST data where available. Deprojected radii were calculated using inclination and position angle data from HyperLeda~\citep{Paturel03}, with the central spaxel defined using the WiFeS 'r-band' images. These data, as well as optical diameters are provided in Table 1. In the case of a few of our later-stage mergers we assume an uninclined disk when calculating deprojected radii-these have no PA or Inclination given in Table 1. This provides the most conservative estimate of a gradient in that system while avoiding ambiguities caused by the complex morphologies of late-stage major mergers.

\subsection{Spectral Fitting}
We analyzed every spectrum using an automated fitting routine written in IDL, UHSPECFIT, which is based on the code created for the work in \citet{Zahid11} and is also employed by and described with example fits in~\citet{Rupke10b, Rich10, Rich11}. Our routine fits and subtracts a stellar continuum from each spectrum using population synthesis models from \citet{Gonzalez05} and an IDL routine which fits a linear combination of stellar templates to a galaxy spectrum using the method of \citet{Moustakas06}. 

Lines in the resulting emission spectra are fit using a one or two-component Gaussian, depending on the goodness of fit determined by the routine. All of the emission lines are fit simultaneously using the same gaussian component or components. Both continuum and emission lines were fit using the MPFIT package, which performs a least-squares analysis using the Levenberg-Marquardt algorithm \citep{Markwardt09}. All of the emission line fluxes used in this paper are subject to a minimum signal-to-noise cut of 5.

\section{Analysis}

\subsection{Extinction Correction}
To account for offset in flux between the red and the blue data cubes for each system, we compare the balmer decrement for H$\alpha$, which falls on the red arm, and H$\beta$ on the blue arm to the decrement between H$\beta$ and H$\gamma$ which both fall on the blue arm. The average scaling factors were generally within a few percent of unity, and were used to scale the red fluxes.

Due to the low flux and thus limited spatial coverage in H$\gamma$, we did not use H$\beta$ vs. H$\gamma$ to generate the extinction maps for our final analysis. Final extinction maps were created for each data cube using the \Ha/\Hb decrement and the extinction curve of \citet{Cardelli89}. These maps were then smoothed with a 2-d boxcar size of 3 spaxels to reduce noise in the maps. The extinction map for each system is provided in the appendix. We then deredden all of the measured emission line fluxes for a system using our derived extinction map.

\subsection{Emission-Line Ratios}
We generate emission line ratios from the extinction-corrected line fluxes to diagnose the source of excitation for each spaxel in the systems we analyze.  Emission line ratio maps are a convenient method of tracing the excitation mechanism within galaxies: region of high \NII/\Ha or \SII/\Ha are generally associated with non-HII region photoionization. We subsequently classify each spectrum of every individual spaxel as HII-region like, composite or AGN-like (LINER or Seyfert) using the \NII/\Ha, \SII/\Ha and \OIII/\Hb ratios and the scheme of \citet{Kewley06}. Emission line ratio maps for each system are given in the appendix, and are shown without extinction correction applied. 

To avoid contamination from non-HII region photoionization, we reject spaxels with emission line ratios that lie above the \citet{Kauffmann03} empirical pure star formatiion line in the \NII/\Ha vs. \OIII/\Hb diagram, as well as points that lie beyond the theoretical pure star formation line in the \SII/\Ha vs. \OIII/\Hb diagram of \citet{Kewley06}. To account for errors we allow points that fall within 0.1 dex to the right of the cutoffs, consistent with the work of ~\citet{Rupke10b}. We do not use the \OI/\Ha vs. \OIII/\Hb diagnostic due to the significantly lower S/N in \OI~for most of our systems. 

The spectra that fall into the composite region and AGN classes are contaminated primarily by shock excitation in our systems. This is apparent in several of the emission line ratio maps, where off-nuclear regions show enhanced emission line ratios. In ~\citet{Rich11} we showed that these higher emission-line ratios are reproduced well by new slow-shock models and investigate the star-forming and shocked components separately using kinematic information from our WiFeS spectra. In the case of the galaxies IRAS F10257-4339 and IRAS F01053-1746, we are able to separate the shocked component from the pure HII-region component kinematically using this method as outlined in \citet{Rich11}. For these two systems we apply a velocity dispersion cuts of 65 km/s and 90 km/s respectively as well as line-ratio classifications described above, leaving only pure HII-region emission line ratios for gas-phase metallicity calculation.

\subsection{Abundance Calcuations}
We calculate the oxygen abundance in each spaxel using several strong-line gas-phase metallicity calibrations. We are unable to use the ``direct'' T$_{e}$ method due to the low flux of the \OIII$\lambda4363$ auroral line and the higher metallicities found in our objects. We instead employ both diagnostics empirically calibrated against the T$_{e}$ method and theoretical methods based on the strengths of other measurable emission-line fluxes, adopting the procedures outlined for each method in the summary of ~\citet{Kewley08}. We measure relative abundances within a given diagnostic, avoiding the problems associated with absolute abundance calculations (e.g. ~\citealt{Angel10a,Angel11}). 

We employ the empirical calibration of ~\citealt{PP04} (PP04), which uses the line ratios \NII/\Ha and \OIII/\Hb. Using these line ratios has the benefit of avoiding the necessity of extinction correction-PP04 developed their calibrations for use in higher redshift systems where proper measures of the reddening may not always be possible.

We also calculate metallicities with the theoretical method of ~\citealt{KD02} (KD02), specifically we use the \NII/\OII~line ratio which is insensitive to variations in ionization parameter. Previous work has shown that the KD02 diagnostic shows less RMS dispersion than other diagnostics for this reason (e.g.~\citealt{Kewley08}). When making comparisons with the  \citet{Rupke10b} and \citet{Kewley10} gradient studies, we convert abundances measured with the PP04 diagnostics into the KD02 diagnostic using the prescriptions of ~\citet{Kewley08} for consistency.

We also investigated the $R_{23}$ (\OII$\lambda3727$+\OIII$\lambda\lambda4959,5007$)/\Hb~measures of abundance by employing the calibrations of \citealt{KK04} (KK04), and \citealt{McGaugh91} (M91). The KK04 method corrects for variations in ionization parameter, by using the \OIII/\OII~line ratio in an iterative fashion to calculate both an ionization parameter and a metallicity. Our results using the $R_{23}$ method are consistent with PP04 and KD02 with somewhat increased RMS scatter and are thus excluded from further discussion in this paper. 

\subsection{Metallicity Maps \& Gradients}
We use the derived abundances to create maps of the metallicity for each system in our sample. We then measure the metallicity gradient as a function of deprojected radius using the metallicity values calculated for each spaxel. We calculate our gradients using an unweighted least-squares linear fit, with errors derived using monte-carlo methods, consistent with the work of \citet{Kewley10,Rupke10b}. Only points which are below the HII-region cut are shown on the metallicity maps and gradient plots. This is evident when comparing the line-ratio maps to the metallicity maps: regions of strong line ratios and/or low surface brightness in the line ratio maps are rejected and not shown on the metallicity maps. IRAS F18341-5744 and IRAS F13373+0105 W for instance are dominated by strong shock-like emission over large areas near the nuclei of those systems.

We calculate gradients in both dex/kpc and dex/R$_{25}$, where R$_{25}$ is the effective optical radius, with the latter measure meant to account for differences in physical scale from system to system. In general we are not able to extend our gradients to the same fractional isophotal radius as \citet{Kewley10,Rupke10b}. In some of our systems this is due to the spatially restricted, highly extinguished nature of the line-emitting gas, coupled in some cases with the limited total field of view of our WiFeS mosaics. Our monte-carlo error calculation takes this into account, and we also test the effect smaller spatial sampling may have on our gradient measures by taking systems with spatially well-sampled gradients recalculating new gradients at various limited spatial scales. The net effect when considering only the inner regions (from 0.0 to 0.4 R/R$_{25}$) is the calculation of a slightly steeper gradient than expected. When other portions (e.g. from 0.2 to 0.6 R/R$_{25}$) of the gradient are sampled, the resulting gradient measured is within a few percent of the expected gradient. These results indicate that our conclusions regarding gradient flattening remain relatively unaffected by spatial sampling.

\subsection{Calibration Discrepancies}
We calculate metallicities for each spaxel using different methods in order to investigate any discrepancies in metallicity measurement introduced by errors in our extinction correction, flux measures and from ionization parameter variation. Due to the wide separation in wavelength of the diagnostic emission lines the KD02 calibration is most sensitive to errors in extinction correction while the PP04 diagnostic is essentially immune to reddening effects due to the close proximity of the line ratios employed. On the other hand, the PP04 calibration should be the most sensitive to variations in ionization parameter, while the KD02 diagnostic is essentially insensitive to ionization parameter.

In the case of very dusty LIRGs, the PP04 diagnostic proves the most useful as it avoids using the very blue and thus very extinguished \OII~line. Indeed the PP04 diagnostic provides the least RMS scatter from a straight line fit in our gradient measures, owing most likely to errors in our extinction correction which increase the scatter in O/H values calculated using KD02. The inability of the PP04 diagnostic to account for variations in ionization parameter, however, is a danger if the parameter varies in any systematic way, especially with radius. We find a correlation between ionization parameter, as measured using the KK04 method, and the difference between metallicities measured using the PP04 diagnostic and the KD02 or KK04 diagnostic. Fortunately, the effect of varying ionization parameter is minor and does not affect the conclusions in this paper.  Overall, the discrepancies between the metallicity calibrations in our analysis are consistent with the comprehensive study of \citet{Kewley08}.

It is also worthwhile to note that our gradients are derived using integrated spectra containing all of the light within each spaxel, including any diffuse emission not associated with HII regions. This is in contrast to the targeted HII-region abundance gradients used in \citet{Kewley10,Rupke10b}. Although we carefully remove non-photoionized spectra, we might still expect a diffuse contribution to line-ratios used in gradient calculation \citep{Alonso09,Alonso10}. Studies have shown, however, that abundances inferred from an integrated spectrum are consistent with HII-region abundances, regardless of the abundance calibration used \citep{Moustakas06b,Rosales11}. \citet{Rosales11} in particular present highly resolved IFU spectroscopy of nearby NGC 628 and show that the abundance gradient derived from their total integrated spectra is very consistent with the gradient derived with spectra isolated from HII-regions. 

\section{Abundance Gradients}
\citet{Kewley10} and \citet{Rupke10b} found a flattening in the metallicity gradients of widely separated pairs of galaxies. Their sample represents systems caught in the earliest stages of merging between first and second pericenter. As the galaxies continue to merge, gas should continue to flow towards the individual galaxy nuclei, fueling the ongoing nuclear starburst and quenching star formation in the tidal remnants. The nuclear metallicity would first see a depression due to the infalling gas, followed by a subsequent enrichment and further depression as the merger progresses to coalescence. The metallicity gradient is continuously flattened by the combined effects of infalling low-metal gas and tidal stretching of the spiral arms and outer portions of each system. In this section we show that our results indicate the trend of flattening metallicity gradient may extend to later merger stages.

\begin{figure*}
\centering
{\includegraphics[scale=0.55]{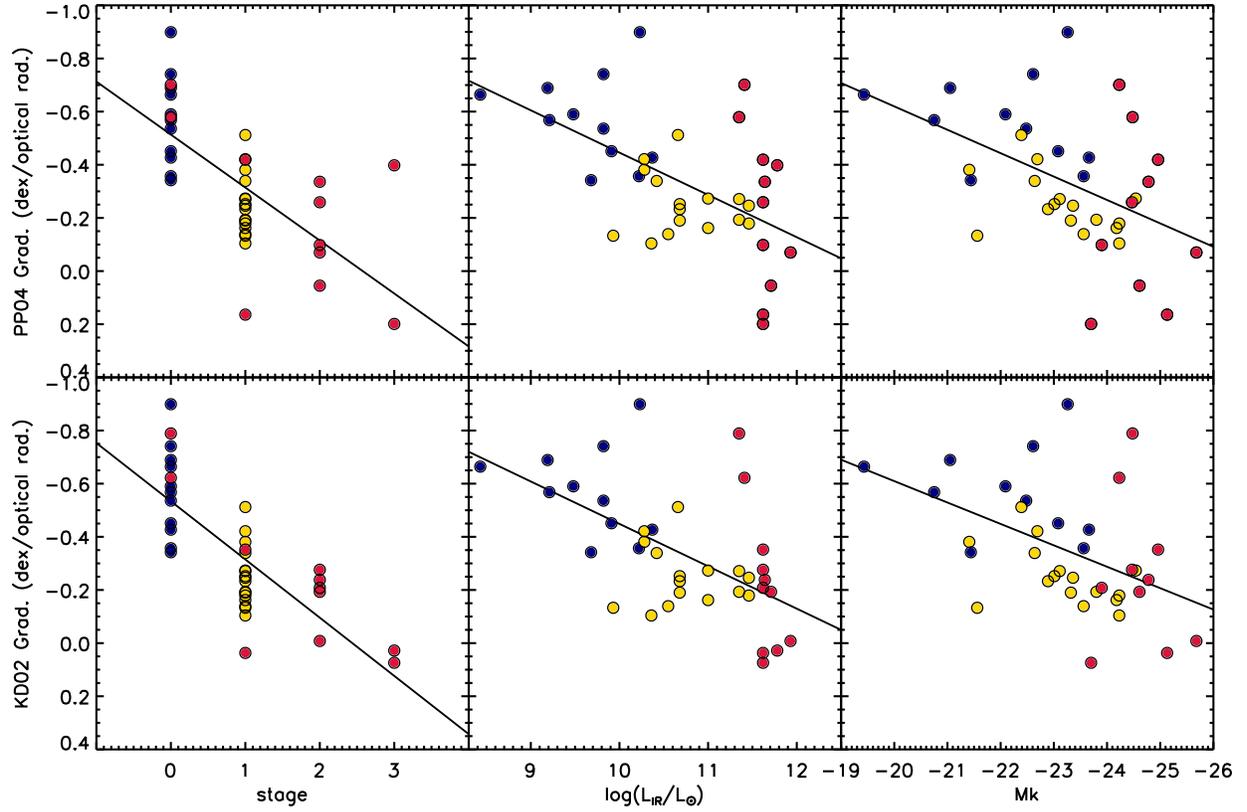}}
\caption{Metallicity gradient as a function of merger stage 0:isolated, 1:widely separated pairs ('a'), 2:close pairs ('b'), 3:coalesced systems ('cde'). The isolated control sample of ~\citet{Rupke10b} is represented by the blue circles, the interacting sample of ~\citet{Kewley10,Rupke10b} by yellow and our data is plotted as red circles. Simple linear regressions are overplotted for each pair of quantities. }
\end{figure*}

\begin{figure*}
\centering
{\includegraphics[scale=0.55]{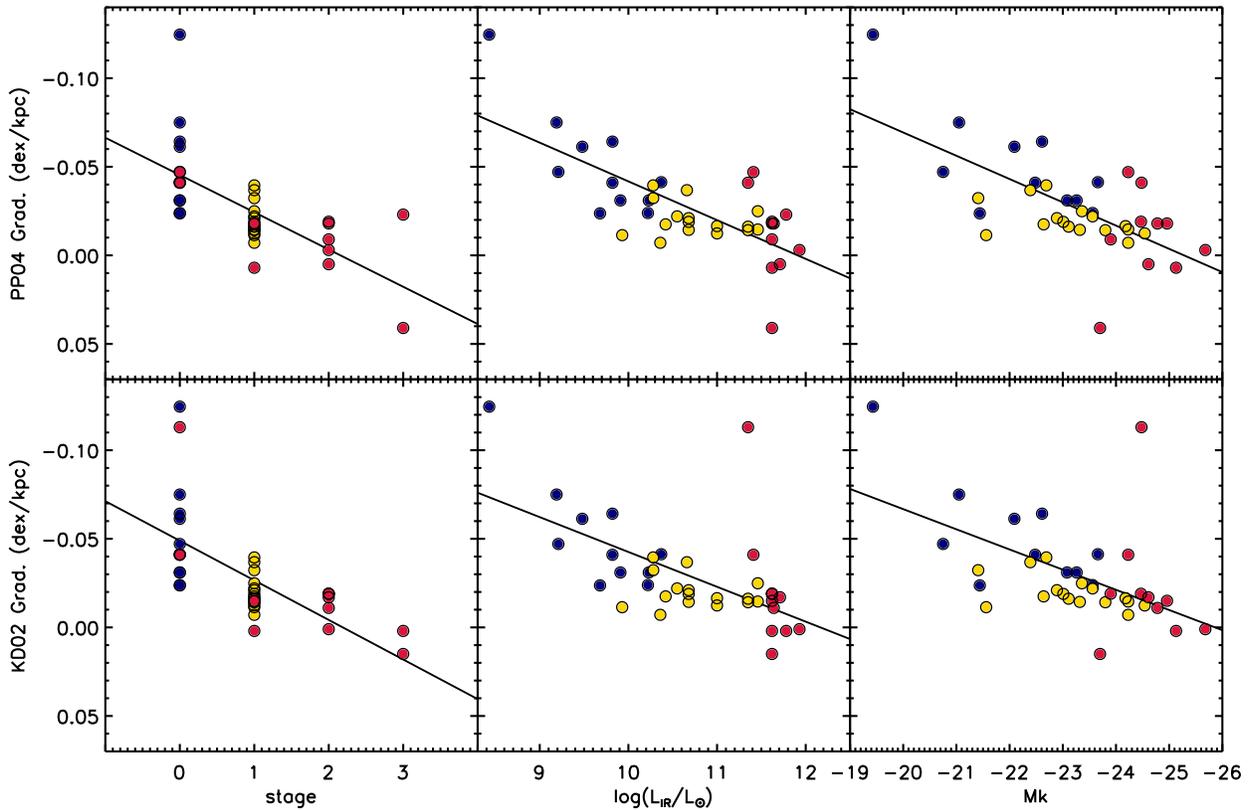}}
\caption{Same as figure 1, but plotting dex/kpc instead of dex/optical radius (R$_{25}$).}
\end{figure*}

\subsection{Merger Stage and Gradient}
In the context of the ~\citet{Yuan10} merger scheme, the control and interacting samples of \citet{Kewley10} and \citet{Rupke10b} are respectively 'isolated' and widely-separated 'a' stage systems, while the majority of the galaxies in our sample are close-pairs in the 'b' stage, further along in the merger process. Our sample also has 3 late-stage coalesced systems ('cde'). It is not clear that all of the stage 'b' LIRGs in our sample will become true ULIRGs given their morphology and L$_{IR}$. ~\citet{Yuan10} suggest that systems with mass ratios closer to unity are more likely to become true ULIRGs during the later stages of the merger process. This could account for systems such as IRAS F21330-3846, which has a comparatively low total L$_{IR}$ despite the fact that it is a more advanced merger. The gas fraction and dust mass of each system prior to the merger also strongly affect the total IR luminosity.

We plot metallicity gradient as a function of merger stage in Figures 1 and 2. Although we do not have a large control sample of isolated LIRGs,  the metallicity gradients of our 2 isolated LIRG systems fall within the range of gradients of the control sample of local, isolated spirals from ~\citet{Rupke10b}. Figure 1 shows that metallicity gradient flattens through each merger stage, with the three coalesced systems showing the flattest gradients. The average gradient in our progressed mergers is similar to the average gradient measured in the widely separated pairs sample of ~\citet{Kewley10,Rupke10b}. Figure 2 shows that our results are unchanged when when metallicity gradient is plotted as a function of optical radius and when it is plotted as a function of absolute deprojected radius in kpc.

Interestingly, the galaxy IRAS F13373+0015 may be showing gradient-flattening effects due to both gas infall as well as the stretching of its spiral arms due to tidal effects from the merger as evidenced by an apparent kink in the metallicity gradient. Unfortunately, the nuclear region is strongly contaminated by shock-like emission, so we excluded gas-phase metallicity measurements from our plot for these radii. If we ignore this contamination and calculate an abundance for the center-most region, the measured nuclear [O/H] does appear depressed with respect to the observed and calculated gradient, which would create a gradient that shows a slightly steeper portion between two flatter regions.

\begin{deluxetable}{ccc} 
\centering
\tablewidth{0pc}
\tablecolumns{3} 
\tablenum{2} 
\tablehead{ \colhead{Variables}  &  \colhead{Correlation Coefficient}  & \colhead{Probability (\%)} }

\centering

\startdata
M$_{K}$, PP04, R$_{25}$ & -0.49 & 0.2\\
M$_{K}$, KD02, R$_{25}$ & -0.48& 0.3\\
L$_{IR}$, PP04, R$_{25}$ & ~0.56 & 0.02 \\
L$_{IR}$, KD02, R$_{25}$ & ~0.61 & 0.005\\
M$_{K}$, PP04, R$_{kpc}$ & -0.60 & 0.008\\
M$_{K}$, KD02, R$_{kpc}$ & -0.59 & 0.01\\
L$_{IR}$, PP04, R$_{kpc}$ & ~0.65 & 0.001\\
L$_{IR}$, KD02, R$_{kpc}$ & ~0.66 & 0.0006

\enddata

\tablecomments{Spearman rank correlation coefficient. Two-sided probability of finding correlation coefficient by chance.} 

\end{deluxetable} 

\subsection{Luminosity and Gradient}
We also plot both L$_{IR}$ and M$_{Ks}$, quantities which should be correlated with merger stage, as a function of metallicity gradient in Figures 1 and 2. L$_{IR}$ traces star formation rate and is correlated with merger stage, while M$_{Ks}$ is unaffected by the high extinctions encountered in LIRGs and is a function of stellar mass (though see sec. 2.2). Values for L$_{IR}$ are taken from \citet{Armus09}, which adjusts the original RBGS \citep{Sanders03} values to the 5-year WMAP cosmological values. M$_{Ks}$ is derived using the distances from \citet{Armus09} and the 2mass 20 mag/arcsec$^{2}$ isophotal k-band magnitudes (consistent with \citet{Rupke10b}).  

~\citet{Rupke10b} concluded there was no strong evidence of a correlation between gradient and either L$_{IR}$ or M$_{Ks}$, though they note that ~\citet{Zaritsky94} pointed to possible correlations when considering dex/kpc. When our observations are combined with the work of ~\citet{Rupke10b}, the case for a correlation between gradient and both L$_{IR}$ and M$_{Ks}$ correlations is strengthened. The Spearman rank correlation coefficients indicate a moderately strong correlation in all cases, though L$_{IR}$ is more strongly correlated than M$_{Ks}$. 

Gradients calculated in dex/kpc are more tightly correlated than those measured in dex/R$_{25}$ (Table 2), consistent with ~\citet{Zaritsky94}.  Increased scatter when considering optical radius may be due in part to the increasingly disturbed morphologies of merging systems. As a merger progresses R$_{25}$ represents a less consistently comparable value as tidal features and the coalescence of both galaxies alter this measure such that it is no longer consistent with R$_{25}$ measured in pre-merger spiral galaxies.

\begin{figure}
\centering
{\includegraphics[scale=0.60]{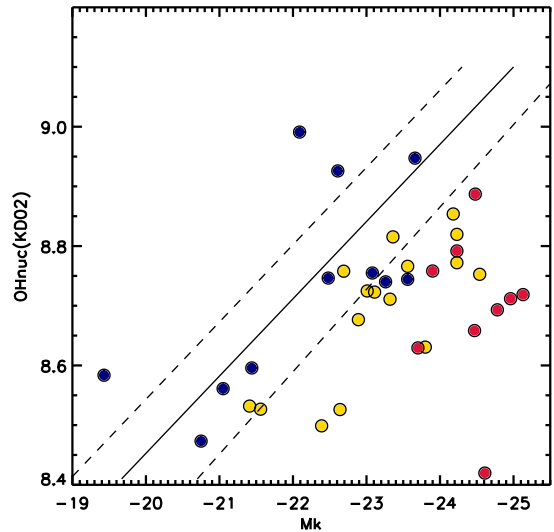}}
\caption{This figure shows the scaled Luminosity \& Nuclear abundance ($L-Z$) relation as provided in Fig. 11 of \citet{Rupke10b}. Our work extends the trend towards lowered nuclear metallicity for higher galactic luminosity. This is consistent with the depressed nuclear metallicity which is expected to be seen in merging systems. The data of \citet{Rupke10b} is plotted in blue (isolated) and yellow (wide pairs) while our LIRGs are plotted in red. Nuclear metallicity is measured by extrapolating the gradient to 0.1R$_{25}$ and scaling to match the approximate L-Z relation of \citet{Salzer05} per the method of \citet{Rupke10b}. }
\end{figure}

\subsection{Nuclear Abundance and Luminosity}
To further compare our sample to previous work, we consider the nuclear abundances from our sample. In order to directly compare to \citet{Rupke10b} we also consider a Luminosity-Metallicity ($L-Z$) relation, rather than a Mass-Metallicity ($M-Z$) relation, as plotted in fig. 3.  We consider the the near-IR L-Z relation of \citet{Salzer05} for consistency with \citet{Rupke10b}. We adopt the same method as \citet{Rupke10b} for nuclear abundance determination: we assign an abundance by extrapolating the calculated gradients to a fiducial radius of 0.1 R$_{25}$. We also adopt the same offset in absolute metallicity with respect to the \citet{Salzer05} relation. Our nuclear metallicities are indeed lower than the expected L-Z relation, again extending previous work on merging systems to higher NIR luminosities.  This result is consistent with previous observations pairs \citep{Kewley06b,Rupke08,Ellison08,Michel08,Peeples09,Sol10} as well as the merging galaxy models we consider in our discussion of gradient evolution with merger stage and properties.

\section{Discussion}
\begin{figure}
\centering
{\includegraphics[scale=0.35]{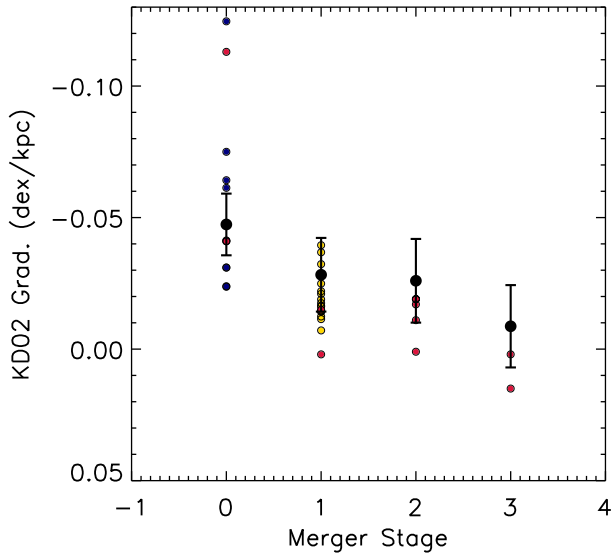}}
\caption{Merger stage vs metallicity gradient as measured from our theoretical merger models. The gradients measured from the merger models in dex/kpc show values and flattening consistent with observations. The average value for the models generated is given by the large black point, the error bars represent the spread in model values. Our gradient data from figure 2 is overplotted for reference, data from \citet{Rupke10b} is overplotted in blue points (isolated) and yellow points (wide pairs) and our LIRG data is overplotted with red points.}
\end{figure}

Our observations are a direct measure of the redistribution of gas within strongly interacting systems. The observational reality of metallicity gradients in non-interacting galaxies is well established ~\citep{Zaritsky94,Vanzee98}. Observations of a change in an established abundance gradient and systematically lower nuclear metalllicities are clear indicators of gas flows and have been seen not only in strong interactions but also in bar structures (e.g. ~\citealt{Vilacostas92,Dutil99,Martin95}). While ~\citet{Friedli94} demonstrated the redistribution of gas traced by metallicity in barred galaxies, only recently have numerical simulations been used to address the detailed physical motions of gas and metals within interacting galaxies.

~\citet{Rupke10a}, ~\citet{Montuori10} and ~\citet{Torrey11} presented the first attempts at using hydrodynamical simulations to address the effects of major galaxy mergers on nuclear abundances and metallicity gradients. The three separate investigations all involve mergers of massive spiral disk galaxies with 1:1 mass ratios. As noted by ~\citet{Rupke10a}, these conditions are typical for the ULIRG formation scenario and are thus a useful direct comparison. 

The models of ~\citet{Rupke10a} represent the simplest set of simulations analyzed for metallicity gradient evolution both in merger parameter space and in model sophistication. Their models use the methodology described in \citet{Barnes04}, but without any ongoing star formation and thus with no chemical enrichment.  This approach clearly captures the effect that low metallicity gas inflows will have on an evolving nuclear metallicity.  \citet{Rupke10a} conclude that the nuclear metallicity can be substantially depressed by gas inflows and that metallicity gradient can be dramatically flattened following tidal tail formation.  They find, not surprisingly, that the depression in the nuclear metallicity is correlated with the gas mass that has migrated to the nuclear region.

~\citet{Montuori10} employ simulations from the GalMer database, described in \citep{Chilingarian10}.  Their simulations include chemical enrichment due to ongoing star formation.  As a result of including chemical enrichment,~\citet{Montuori10} find that the nuclear metallicity changes non-monotonically as a function of the merger as chemical enrichment partially offsets the effect of low-metallicity gas inflows.  

~\citet{Torrey11} have studied merger induced nuclear metallicity evolution while considering star formation and chemical enrichment over a relatively wide range of merger parameter space.  The analysis in~\citet{Torrey11} explicitly differentiated the effects of initial metallicity gradient re-distribution from ongoing metal enrichment on the nuclear metallicity evolution -- allowing for a basic template for merger nuclear metallicity evolution to be presented.  The simulations of~\citet{Torrey11} included systems with varied initial mass and several merger orientations -- allowing for a more complete comparison to observations.


\begin{figure*}
\centering
{\includegraphics[scale=0.40]{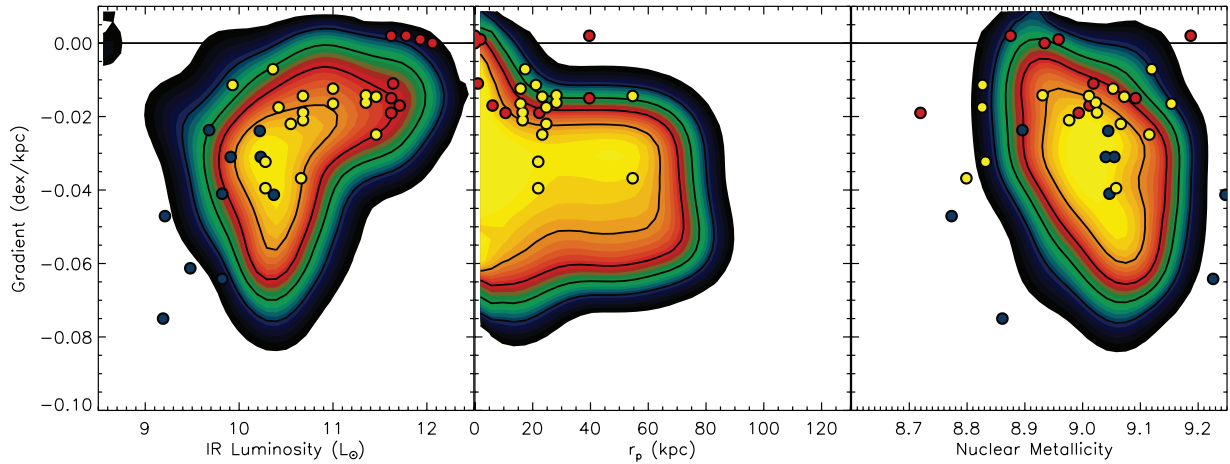}}
\caption{Metallicity gradient compared to L$_{IR}$, nuclear separation and nuclear metallicity. The color contour plots are generated using measurements from the merger models while data from \citet{Rupke10b} is overplotted in blue points (isolated) and yellow points (wide pairs) and our LIRG data is overplotted with red points. These plots show good agreement between models and observations while providing information about how these quantities roughly track with merger progress. L$_{IR}$ derived from the simulations is calculated based on the instantaneous star formation rate using the formulae of \citet{Kennicutt09}.}
\end{figure*}

\subsection{New Models}

\begin{figure}
\centering
{\includegraphics[scale=0.40]{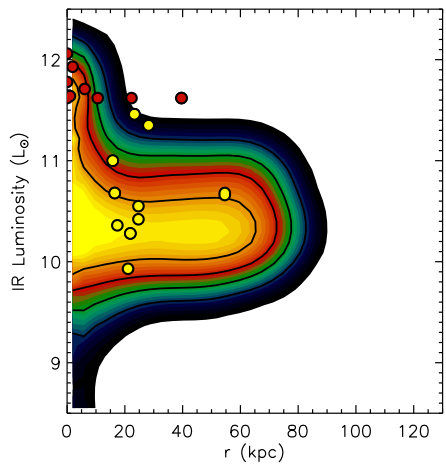}}
\caption{L$_{IR}$ vs. projected separation measured from the merger models as plotted in fig. 5 with our data overplotted as red points and the widely separated systems of \citet{Rupke10b} overplotted in yellow. Despite the agreement between observations and models this plot shows projected separation as a less helpful measure of merger progress. }
\end{figure}
The interpretation of the data presented in this paper can benefit by considering a realistic theoretical model for the evolution of interacting galaxies.  In principle, one could use numerical models to directly simulate the merging systems that have been presented in this paper.  However, this would require accurately setting a large number of parameters for each merging system (e.g., the initial stellar mass of each galaxy, the initial gas fraction, the initial morphology, the merging orbital orientation, etc.).  Fortunately there are certain aspects of major mergers, (e.g. the formation of tidal features) which are generically associated with strong tidal encounters and do not sensitively depend on the detailed characteristics (e.g. the gas fraction) of the merging galaxies.  

In the context of metallicity gradient evolution, we can learn a lot about the expected evolution of a population of merging galaxies by exploring a limited number of carefully selected merger simulations.   Our goal is to present a realistic theoretical model to augment the interpretation of our observational data.  To achieve this, we build a limited merger simulation suite that we can explore and understand in detail.  Although this approach will leave certain areas of merger parameters space unexplored, we still expect the characteristics and tends present in the simulations to match that of the observations which allows us to physically probe the driving forces behind the observed evolving metallicity gradients presented in this paper.

We compare our observations to a suite of galaxy merger simulations (\citealt{Torrey11}) carried out using GADGET-2~\citep{Springel05}.  Our simulated merger suite consists of 32 simulations which are achieved by using 16 merger orientations (orientations a-p in \citet{Torrey11}) with 2 progenitor galaxies.  The two progenitor galaxies for our simulations both have equal total system mass, but with slightly different initial disk profiles.  Both systems include a~\citet{Hernquist90} dark matter halo ($M_H = 5.1 \times 10^{11} M_\odot$, $a = 22$ kpc) with rotationally supported exponential gas and stellar disks ($R_{d,1} = 3.2$ kpc, $R_{d,2} = 4.6$ kpc).  The initial gas fraction of the system is 25\%, which will decrease with time due to star formation.  

A substantial discussion of the kind of models used for comparison in this paper as well as further analysis of theory derived from those models can be found in \citet{Torrey11}.  Here, we briefly describe our methods for tracking the nuclear metallicity and metallicity gradient, which serves as the key component of our simulation analysis.   Our simulations track a single metallicity scalar value that is allowed to increase as a result of chemical enrichment from star formation.  To compare our simulations to the observations, we focus only on the Oxygen abundances, which we assume to be 30\% of the total metal mass.   Because most Oxygen is produced in type II supernovae which return their mass and metals to the ISM over relatively short (i.e. $\sim10^7$ year) timescales, we assume instantaneous metal enrichment which we practically incorporate into our simulation by setting the rate of metal production proportional to the local star formation rate.

We impose metallicity properties on our galaxies at the beginning of the simulation such that they conform with observations of galaxies in the local universe.  Namely, we initialize the nuclear metallicity of our galaxies onto the sloan digital sky survey derived mass-metallicity relation~\citep{Tremonti04} with the radial metallicity gradient following a decaying exponential profile with a characteristic length proportional to the disk scale length~\citep{Zaritsky94}.

Using this model, the metallicity in our galaxy is determined by the re-distribution of the initial metallicity gradient due to bulk movements of gas induced by the strong gravitational interaction and chemical enrichment associated with starburst driven star formation.  At any time in the simulation, the metallicity can be calculated for a region of space by finding the mass weighted average metallicity of all gas particles in the specified volume.  For example, we calculate the nuclear metallicity by finding the mass weighted average metallicity for all gas particles within a central 1kpc sphere of the galaxies center (defined by the potential minimum).  Or, to determine the metallicity gradient, we calculate the average metallicity within a series of concentric spherical shells centered on the galaxies potential minimum, and find the best linear fit of the resulting radius-metallicity data points.  

For an instructive comparison to the observed systems presented in this paper, we track the nuclear separation, nuclear metallicity, metallicity gradient, and IR Luminosity during the merger.  Using the data from all simulations, we can then form an expected evolution and distribution of galaxy properties during the merger sequence.  The results are plotted in figures 4, 5 \& 6. 

\subsection{Merger Progress Tracers}
In Figure 4 the merger model metallicity gradient is plotted as a function of merger stage where the simulated mergers are classified using the same formalism as our observational analysis.  The solid black circles show the mean simulated gradient value of all merging systems at a given merger stage with the error bars representing the 1-$\sigma$ standard deviations derived from the spread in gradient values for the simulated galaxies in that bin.  By inspection, we find that both the models and observations show a clear trend toward shallow metallicity gradients with increasing merger stage.  In the models we find a strong change in the gradient that occurs immediately after first pericentric passage, when the formation of tidal features stretches the initial metallicity gradient and gas inflows begin to flood the nuclear region with low metallicity gas.  The gradient flattening continues as the merger progresses and by merger stage cde (i.e. coalesced systems), both the models and observed systems have metallicity gradients consistent with being flat.   

Figure 5 shows the evolution of the metallicity gradient as a function of IR Luminosity, projected separation, and nuclear metallicity. L$_{IR}$ derived from the simulations is calculated based on the instantaneous star formation rate using the formulae of \citet{Kennicutt09}. Each panel shows the distribution of simulated galaxies---as the background colored contours---along with the observational data.  The colored contours represent the local density of simulated data points (where each snapshot from each merger counts equally as one data point) with the solid black lines denoting regions that contain 50, 70, 80, and 90\% of the simulated data points. As discussed above, we find that the gradient becomes increasingly shallow as the merger progresses.  We also find the IR Luminosity increases and the nuclear metallicity becomes slightly depressed. This is not unexpected, as the same fundamental process is driving the evolution of all of these quantities. Strong tidal encounters that are associated with the merger sequence will form tidal features that will stretch or flatten the metallicity gradient. The same tidal encounters are responsible for driving gas into the central region, which lowers the central metallicity and enhances starburst activity. Therefore, we expect that the association of flat metallicity gradients with enhanced IR luminosity and depressed nuclear metallicity is a robust byproduct of the merger process.

Figures 5 and 6 show the metallicity gradient and nuclear metallicity against observationally projected separation. Again we find that the observations and merger models remain in general agreement, but the projected separation is not as useful for tracing merger progress. We find the intermediate and late stage mergers are closely clustered in projected separation. This is a result of the non-monotonic evolution of the nuclear separation with merger stage. Statistically analyzing large surveys should indeed show enhancements in the star formation rate, depressions in the nuclear metallicity, and a general flattening of the metallicity gradient that evolves with projected separation. However, separating intermediate and late stage mergers cannot effectively be carried out using only the projected separation information.

\subsection{Further Modeling Considerations}
These conclusions will remain fundamentally unchanged if we applied a more extensive merger simulation suite based on the physical justification that the formation of tidal features and driving of gas to the galactic central region -- the physical engines behind the observed trends presented in this paper -- are events that are generically associated with the merger process.  The formation of tidal tails in particular is a well known result of a tidal encounter~\citep{Toomre72}, which can depend on the merger orbital parameters~\citep[e.g.][]{Donghia10}.  If the tidal encounter is sufficiently weak because the perturbing galaxy is not massive enough or too distant, then we will not expect strong tidal features to form and we would not expect to see a dramatic flattening of the metallicity gradient.  However, in the regime of strong tidal encounters, we expect tidal tails to form and the metallicity gradient to flatten without any sensitive dependencies on, e.g., varied gas fractions or the presence of a bulge.

The same arguments apply when considering the driving of gas into a galaxy's central region.  The tidal encounter sets up mis-aligned bars in the stellar and gaseous components allowing the gas to lose angular momentum and fall into the nuclear region~\citep{Barnes96}.  Additional considerations do exist, such as the stabilizing effect of stellar bulges against axis-symmetric perturbations during distant tidal encounters~\citep{Mihos96} or the reduced torquing efficiency of very gas rich systems~\citep{Hopkins09}.  However, neither of these effects will prevent a strong gravitational interaction from driving gas inflows leading into final coalescence~\citep[see, e.g.,][for detailed justifications]{Mihos96,Hopkins09}.  

These comparisons provide a useful measure of the accuracy of merger models and strengthen the use of such models to make predictions about various aspects of galaxy mergers and their effect on galaxy evolution. A more thorough discussion of the kind of models used for comparison in this paper as well as further analysis of theory derived from those models can be found in \citet{Torrey11}.

When our new observations are combined with previous work, there is also evidence of a correlation between metallicity gradient and both L$_{IR}$ and M$_{K}$ , quantities that loosely trace merger progress. Assigning a precise merger stage is non-trivial. Determining a rough estimate of the time elapsed since first pericentric passage for instance is not possible when considering only the morphology and projected separation of a merging pair. One solution could be the use of software like Identikit~\citep{Barnes09,Barnes11} to constrain merger parameters using observed velocity information. A more reliable merger stage could be assigned once an accurate model is obtained, allowing for a more detailed study of the change in various physical quantities as a merger progresses.

\subsection{Presence of Shocks}
There are other factors which hamper gradient measurements as mergers progress: ~\citet{Yuan10} note a marked increase in AGN/LINER influence on optical spectra as a function of L$_{IR}$ and merger stage, an effect also seen by ~\citet{Armus90,Veilleux95,Goto05,Ellison11}. This effect, coupled with the increasingly compact region of star formation as the remaining gas in a merger is concentrated entirely within smaller nuclear regions makes measurement of spatially resolved gas-phase metallicities difficult and prone to contamination from an AGN. In addition, increased contamination from radiative shocks would be expected in the most IR luminous systems, consistent with increasing gas inflows and the observed increase in supergalactic winds as a function of L$_{IR}$ ~\citep{Rupke05}. In our overall IFU sample of 27 U/LIRGs, $\sim$2/3 of the systems show a contribution from extended shock excitation, fractionally increasing as the merger progresses. IFU observations of sufficiently high spatial and spectral resolution should be able to overcome these issues as long as there is of measurable star formation.

\section{Summary and Conclusions}
We have measured the gas-phase oxygen abundances in 9 nearby LIRG systems using integral field spectroscopy. We measured emission line fluxes for all of our IFS data cubes and used this information to create maps of excitation and metallicity measured with various strong-line metallicity diagnostics. We calculated radial abundance gradients using our metallicity maps coupled with radii and deprojected radii calculated for each system. Our results agree with recent observations of gradient flattening in interacting pairs ~\citep{Kewley10,Rupke10b} and indicate some evidence of flattening at later stages of the merger process, though further observations are needed to confirm this result.

Our results are consistent with recent numerical models of metallicity and gas flows in major mergers ~\citep{Rupke10a,Montuori10,Torrey11}. We compare our observations with a new set of merger models used to track various quantities including metallicity gradient, nuclear metallicity, L$_{IR}$ and separation. These new models track gradients and enrichment carefully throughout the merger process, including the effects of ongoing chemical enrichement for the first time. Our observations agree remarkably well with the models, adding observational support for theoretical predictions about gas flows in numerical merger models. Our comparison between model and observation also indicate the efficacy of metallicity gradient and L$_{IR}$ in tracking merger progress, while showing nuclear separation as less useful. 

The latest stages of a major merger event prove the most difficult period to analyze and interpret using our observational methods and data. Investigating the gradient evolution using numerical simulations of post-merger systems requires fully cosmological simulations where accretion of prestine gas at late times can help reestablish an abundance gradient \citep{Torrey11}. Investigation of this phenomenon also requires a finer spatial scale and deeper observations than we have achieved with our current data set. An increasing contribution from radiative shocks and AGN to the emission line spectra of late stage mergers may inhibit these measurements, necessitating non emission-line abundance diagnostics. 

\begin{acknowledgements}
We would like to thank the referee for his/her helpful comments, which helped clarify many points in this paper. Dopita, Kewley and Rich acknowledge ARC support under Discovery  project DP0984657. This research has made use of the NASA/IPAC Extragalactic Database (NED) which is operated by  the Jet Propulsion Laboratory, California Institute of Technology, under contract with the National  Aeronautics and Space Administration. 

\end{acknowledgements}

\bibliographystyle{apj}

\pagebreak

\appendix

\begin{deluxetable*}{lccccccc} 
\tablewidth{0pc}
\tablecolumns{7} 
\tablenum{3} 
\tablehead{ \colhead{IRAS \#     }  &  \colhead{$\Delta$(dex/R$_{25}$) PP04}  & \colhead{$\Delta$(dex/R$_{25}$) N2O2} & \colhead{$\Delta$(dex/kpc) PP04} & \colhead{$\Delta$(dex/kpc) N2O2} & \colhead{Int. PP04} & \colhead{Int. N2O2} \\ 
\colhead{(1)} & \colhead{(2)} & \colhead{(3)} & \colhead{(4)} & \colhead{(5)} & \colhead{(6)} & \colhead{(7)}  }

\centering
\startdata
F01053$-$1746        &~$0.055\pm0.027$ &  -$0.193\pm0.059$ &  ~$0.005\pm0.003$  &   -$0.017\pm0.005$  & $8.653$ & $8.739$  \\
~ 08355$-$4944       &~$0.199\pm0.049$ & ~$0.074\pm0.086$ &  ~$0.041\pm0.011$  &  ~$0.015\pm0.018$  & $8.791$ & $8.922$   \\
F10038$-$3338        & -$0.398\pm0.136$ & ~$0.028\pm0.204$ &   -$0.023\pm0.008$  &  ~$0.002\pm0.013$  & $9.026$ & $8.958$   \\
F10257$-$4339        & -$0.336\pm0.119$ &  -$0.238\pm0.074$ &   -$0.018\pm0.006$  &   -$0.011\pm0.004$  & $9.032$ & $9.017$   \\
F13373$+$0105 W  & -$0.419\pm0.017$ &  -$0.352\pm0.031$ &   -$0.018\pm0.001$  &   -$0.015\pm0.002$  & $9.037$ & $9.047$   \\
F13373$+$0105 E   &~$0.164\pm0.075$ & ~$0.037\pm0.108$ &  ~$0.007\pm0.004$  &  ~$0.002\pm0.005$  & $8.984$ & $9.015$   \\
F17222$-$5953        & -$0.701\pm0.017$ &  -$0.622\pm0.027$ &   -$0.047\pm0.002$  &   -$0.041\pm0.002$  & $9.133$ & $9.154$   \\
F18093$-$5744 N    & -$0.259\pm0.023$ &  -$0.276\pm0.054$ &   -$0.019\pm0.002$  &   -$0.019\pm0.005$  & $8.971$ & $9.986$   \\
F18093$-$5744 S    & -$0.098\pm0.052$ &  -$0.208\pm0.065$ &   -$0.009\pm0.005$  &   -$0.019\pm0.006$  & $9.045$ & $9.079$   \\
F18341$-$5732        & -$0.579\pm0.165$ &  -$0.789\pm0.193$ &   -$0.041\pm0.013$  &   -$0.113\pm0.015$  & $9.231$ & $9.266$   \\
F19115$-$2124        & -$0.070\pm0.023$ &  -$0.008\pm0.035$ &   -$0.003\pm0.002$  &  ~$0.001\pm0.003$  & $8.955$ & $8.876$   
\enddata

\tablecomments{Derived metallicity gradient slopes and intercepts in dex/kpc and dex/R$_{25}$ for our sample.} 
\end{deluxetable*} 

\section{Notes on Individual Objects}

\subsection{Isolated Systems}
There are 2 isolated systems in our 	sample. These systems have properties most consistent with the~\citet{Rupke10b} control sample, with well established metallicity gradients not yet disturbed by interactions. Both systems are lower luminosity LIRGs, with L$_{IR}\sim10^{11.4}$L$_{\odot}$.

\emph{IRAS F17222-5953} (ESO 138-G027).\textemdash This system is more akin to a typical, non-interacting, strongly starbursting spiral galaxy in our sample. It is in the vicinity of a few other galaxies, including the similarly bright ESO 138-G026, but even the nearest galaxy is at a projected distance of over 100 kpc and IRAS F17222-5953 is not yet interacting with any of these systems. Our IFU data applied to the \NII/\Ha v \OIII/\Hb BPT diagram show a clean curve following the shape of the SDSS sequence of local star forming galaxies (e.g. \citet{Kewley06}), but with a slight apparent shift in total \NII/\Ha. We interpret this shift as an overall Nitrogen enhancement ~\citep{Perez-Montero09}.

\emph{IRAS F18341-5732} (IC 4734) Like IRAS F17222-5953, our second isolated galaxy is in the vicinity of a few other luminous galaxies, but is not yet undergoing any interactions. Our nuclear spectra are dominated by a LINER combined with an aging stellar population. The strongest sights of star formation are where the bar in this galaxy meet the spiral arms evidenced by the two strong clumps of HII-region like points seen in the line-ratio and metallicity. \Ha imaging by \citet{Dopita02} shows further knots of star formation along the spiral arms and our nuclear spectra also show signs of an aging stellar population in the nucleus of IC 4734.

\subsection{Widely Separated Systems}
The interacting sample of ~\citet{Rupke10b} consists of widely separated systems. Our sample has only a single pair of galaxies at this stage of interaction, the interacting system Arp 240.

\emph{IRAS F13373+0105 W} (NGC 5257) Although NGC 5257 and its equal mass partner NGC 5258 are still widely separated and retain much of their structure, they exhibit interaction features including tidal tails and a bridge between the two galaxies. The spiral arms show regions of strong star formation with some signatures of post-starburst populations away from the nucleus. The nuclear regions of NGC 5257 are dominated by an older stellar population as the intense nuclear starburst associated with the later stages of major mergers has not yet begun. The extinction is higher in the nuclear regions and the measurable line-ratios place the nucleus in the composite region of the standard diagnostic diagrams, indicating possible LINER activity-previous nuclear observations and integrated spectrophotometry are consistent with the higher nuclear extinction and overall line ratios we observe \citep{Veilleux95,Kewley01b,Moustakas06}.

\emph{IRAS F13373+0105 E} (NGC 5258) NGC 5258 is a near-twin to NGC 5257 in mass and luminosity. Our spectra indicate higher extinction and a flatter gradient, though the extent of measurable HII-region metallicities is smaller than in NGC 5257. Again similar to NGC 5257 the nuclear region of NGC 5258 is dominated by older stars and very little line-emission and there is some evidence of younger post-starburst populations away from the nucleus. The strongest line emission is associated with the knot of star formation to the southwest of the nuclear region.

\subsection{Closely Interacting Systems}
Our sample has several closely interacting systems, we calculate gradients for 5 of them in this paper. These systems span a range of L$_{IR}$ from typical LIRG luminosities to nearly ULIRG. IRAS F01053-1746, IRAS F10257-4339 and IRAS F19115-2124 are the most indicative of this class, while IRAS F18093-5744 N/S is a triple system with two widely separated spirals and a third, less massive closely interacting component. Although IRAS F01053-1746 and IRAS F10257-4339 are both composed of two individual galaxies, they have progressed along the merger sequence to the point where we are unable to effectively assign gas metallicities to either of the galaxies. We thus treat both IRAS F01053-1746 and IRAS F10257-4339 as individual systems with the brightest optical nucleus defined as the zero point for a metallicity gradient. 

\emph{IRAS F01053-1746} (IC 1623) This system contains two very closely interacting spiral galaxies. It is kinematically very complex and exhibits evidence of widespread radiative shocks, which are analyzed in detail in ~\citet{Rich11}. The deprojected radii are plotted using the hyperleda values for the western, less obscured galaxy IC 1623 A. The eastern system is very intensely star-forming as seen in the infrared (e.g. \citet{Howell10}), but is so enshrouded in dust that our optical spectra do not trace star-forming gas.

\emph{IRAS F10257-4339} (NGC 3256) This advanced merger is quite nearby and well studied. As with IRAS F01053-1746, one of the galaxies in IRAS F10257-4339 is very extinguished. The second system and its nucleus are revealed at longer wavelengths, south of the main optical nucleus \citep{Alonso02,Rothberg10}.  Deprojected radii are calculated using the main optical nucleus. The gas and tidal tails in this system extend quite far (e.g. \citealt{Rothberg10}) our IFU mosaic covers only the central 6 kpc, though this appears to be the physical extent of most of the ongoing star formation in this system. NGC 3256 also shows evidence for widespread shocks in our data and is discussed alongside IC 1623 in \citet{Rich11}.


\emph{IRAS F18093-5744 N} (IC 4687) We classify this galaxy as a close merger; in fact it is a member of a triplet. IC 4687 is undergoing a close merger with the less massive starburst IC 4686, classed as a Wolf-Rayet galaxy by \citet{Kovo99,Fernandes04}. IC 4687 itself is a is also in a wide merger with the equally massive spiral IC 4689. The archived Hubble Space Telescope images of IC 4687 show a complex morphology tangled up with IC 4686: gas and dust from IC 4687 appear to be obscuring the less massive system. Our IFU data covers the entirety of IC 4686/4687 and the metallicities we measure are consistent with the expected metallicties in the outskirts of IC 4687 as extrapolated from the gradient we present in this paper as well as a low-metallicity, flattened gradient in IC 4686. The kinematic information from our data also indicates that we are indeed seeing gas from both systems.

\emph{IRAS F18093-5744 S} (IC 4689) This spiral galaxy is slightly less massive and luminous than IC 4687 and less well-observed, though it is still intensely star forming ~\citep{Howell10}. It is less morphologically disturbed than the other two interacting galaxies IC 4686/4687, though its gradient is quite flattened already according to our observations. Although it is widely separated from IC 4687, we include it as part of the closely interacting system of IRAS F18093-5744.

\emph{IRAS F19115-2124} (ESO 593-IG 008) \citet{Vaisanen08} analyze the morphology of this close pair in detail using Adaptive Optics K-band images. Because of the complex structure we adopt the projected radii for gradient calculations and make no attempt to separate the pair into two separate gradients. Given the flat values for metallicity seen, however, adapting an inclination and/or two disks would only lead to an even shallower calculated slope in the case of this system. 

\subsection{Coalesced systems}
Our metallicity sample has two systems which are classified as coalesced mergers (cde). The most advanced mergers have complex orientations and morphologies which present a difficulty when calculating deprojected radii. As a conservative estimate we simply use the projected radii in our gradient calculations.

\emph{IRAS 08355-4944} 
While Hubble Space Telescope I-band images show remnant tidal tails extending nearly 20 kpc, only the central 5 kpc or so appears to harbor the intense ongoing star formation in this system. Our spectra are dominated by HII regions, with evidence in some regions of a blue-shifted component with low-velocity shock-dominated line ratios which could be associated with a galactic wind.

\emph{IRAS F10038-3338} (ESO 374-IG032)
This post-merger exhibits significant ongoing star formation in its southwestern tidal arm unlike the two other coalesced systems in our sample. The total line emission in this region is much weaker than in the nucleus, inducing a large uncertainty in the extinction map and [OII] lines, creating the discrepant values seen in the metallicity gradient plots for this galaxy. Either gradient, however, is consistent with the overall flattening trend seen in all of our systems. This system hosts an OH megamaser and has soft x-ray emission, all consistent with the advanced stage of merging and increasingly intense nuclear starburst \citep{Henkel90,Staveley-Smith92,Darling02,Iwasawa09}. Our spectra also show evidence for extended off-nuclear shock emission dominating in areas where there is little to no evidence of ongoing star formation, consistent with the IFU observations of \citet{Monreal10}.

\begin{figure*}
\centering
{\includegraphics[scale=0.56]{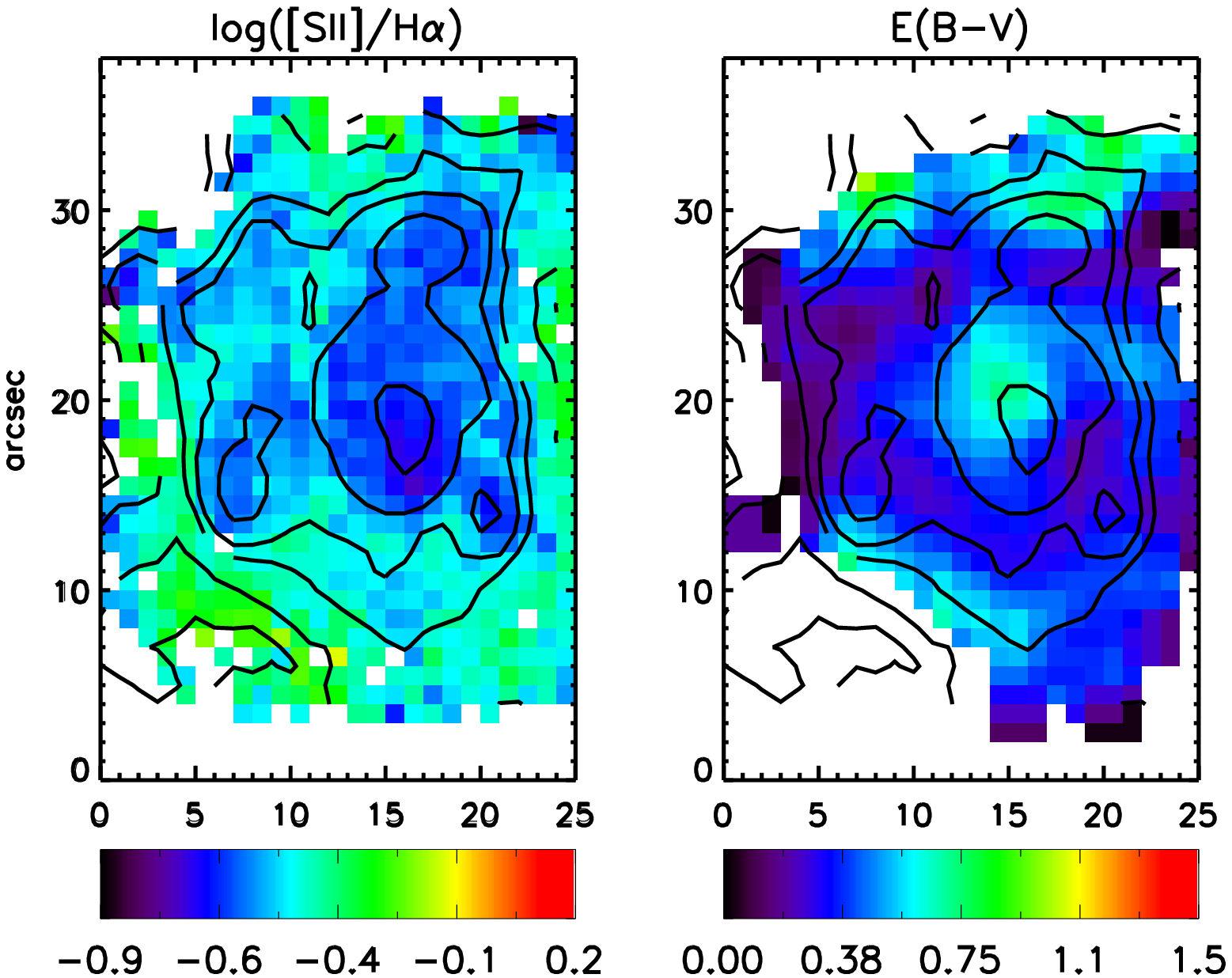}}
{\includegraphics[scale=0.75]{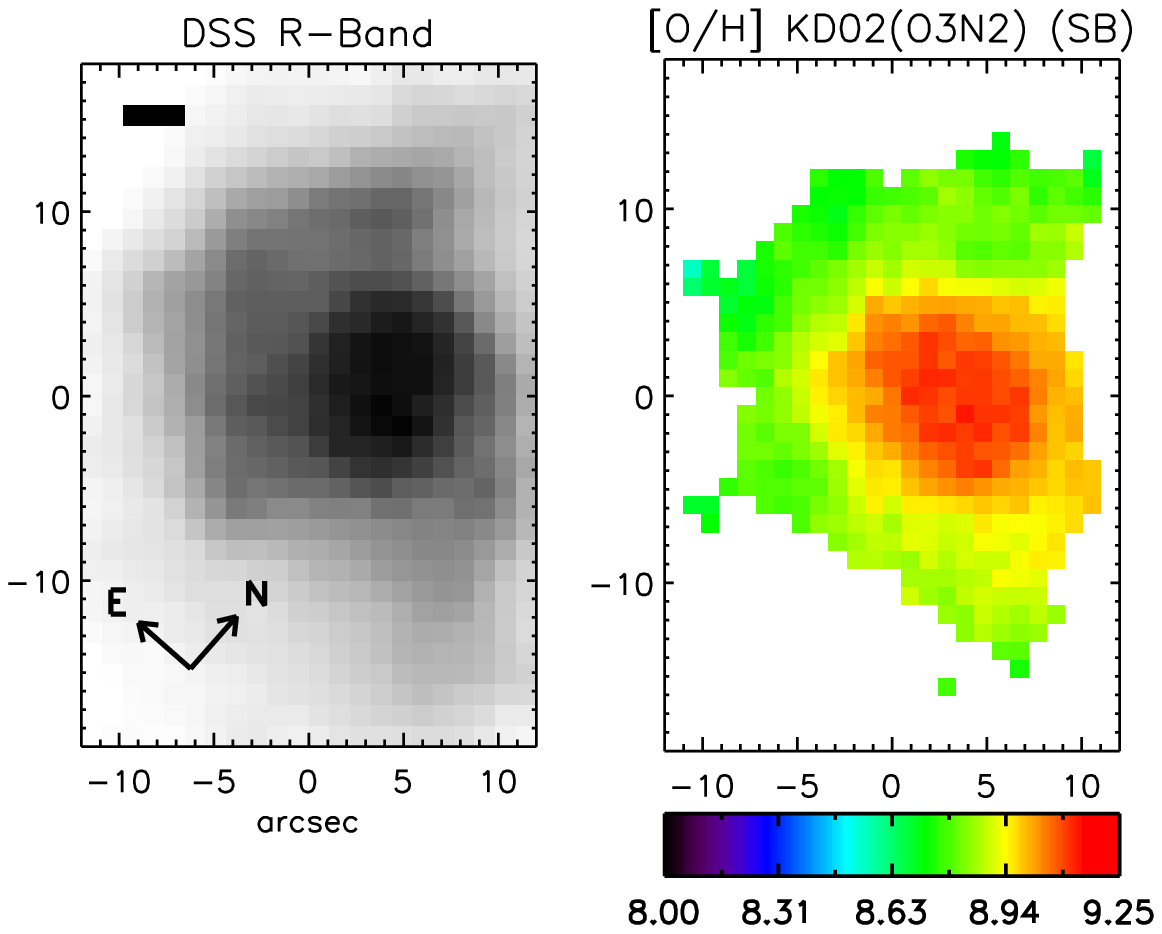}}
{\includegraphics[scale=0.99]{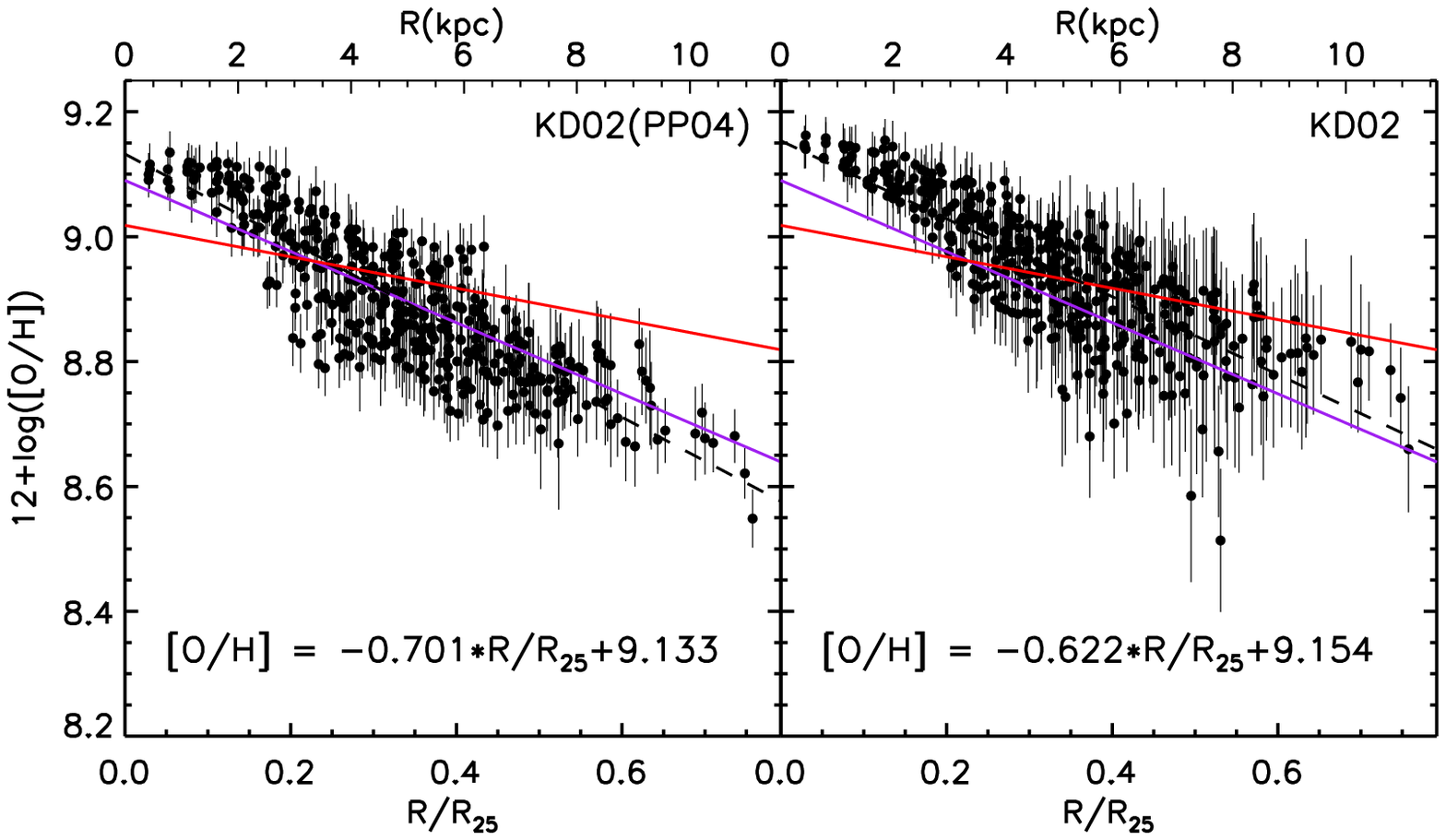}}

\caption{Maps and metallicity gradients for IRAS F17222-5953. The top left two panels show \NII /\Ha \& \SII /\Ha line ratio maps of the system with \Ha contours measured from our data overlaid. The top right panels show an image of the galaxy compared to the metallicity map (derived from the PP04 O3N2 diagnostic, with values scaled to the KD02 diagnostic. The black bar in the system image corresponds to a distance of one kpc. The bottom two panels show the run of abundance with radius calculated from the PP04 (scaled to KD02) and the KD02 diagnostics. The least-squares fit metallicity gradient is overplotted in each case. The average abundance gradients for the isolated (purple) and interacting (red) sample from \citet{Rupke10b} are plotted in the right hand panels for comparison. As described in the text, only points that pass our HII-region cut in the diagnostic diagrams are plotted in the metallicity map and abundance gradients.}
\end{figure*}

\begin{figure*}
\centering
{\includegraphics[scale=0.51]{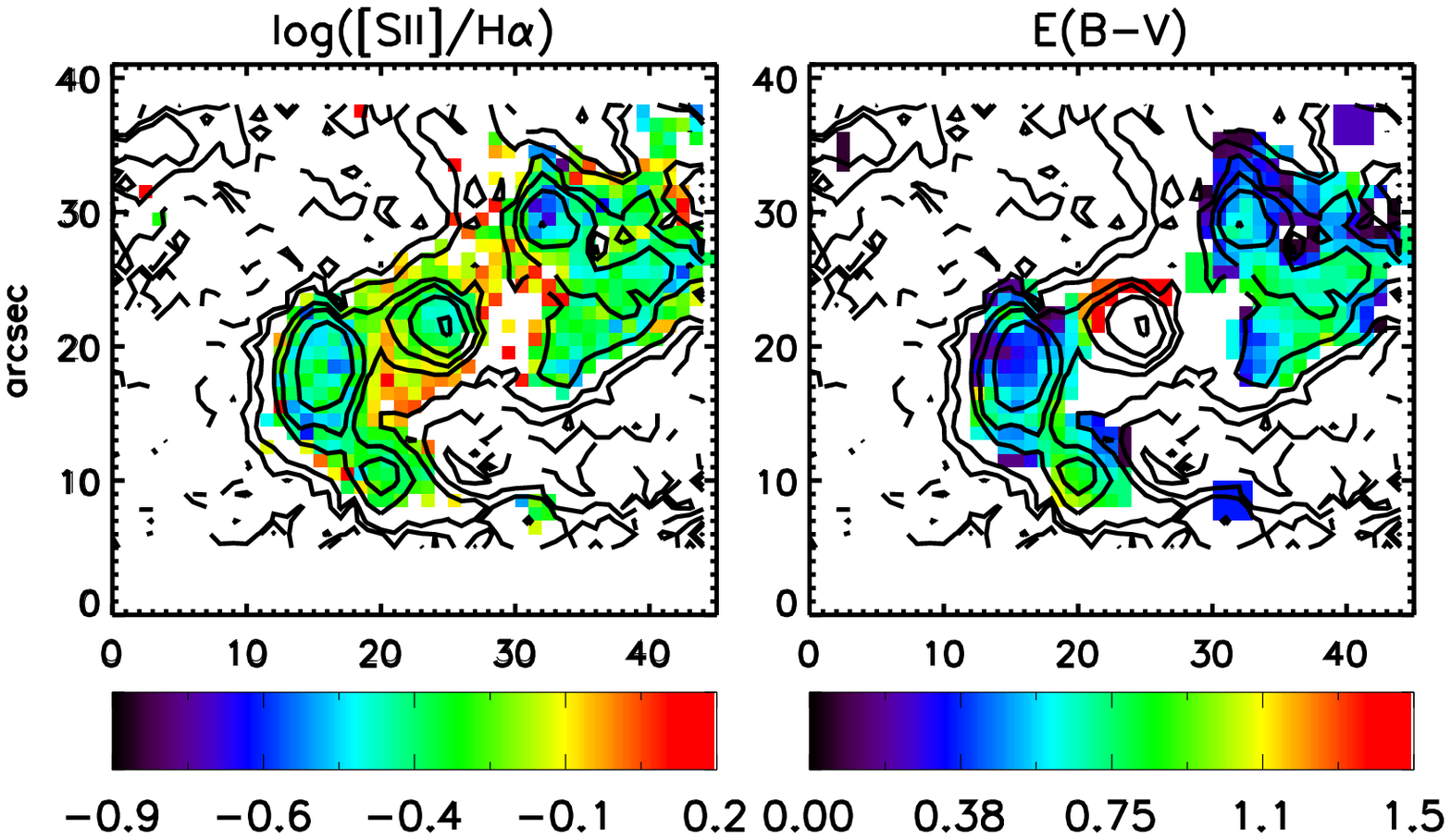}}
{\includegraphics[scale=0.78]{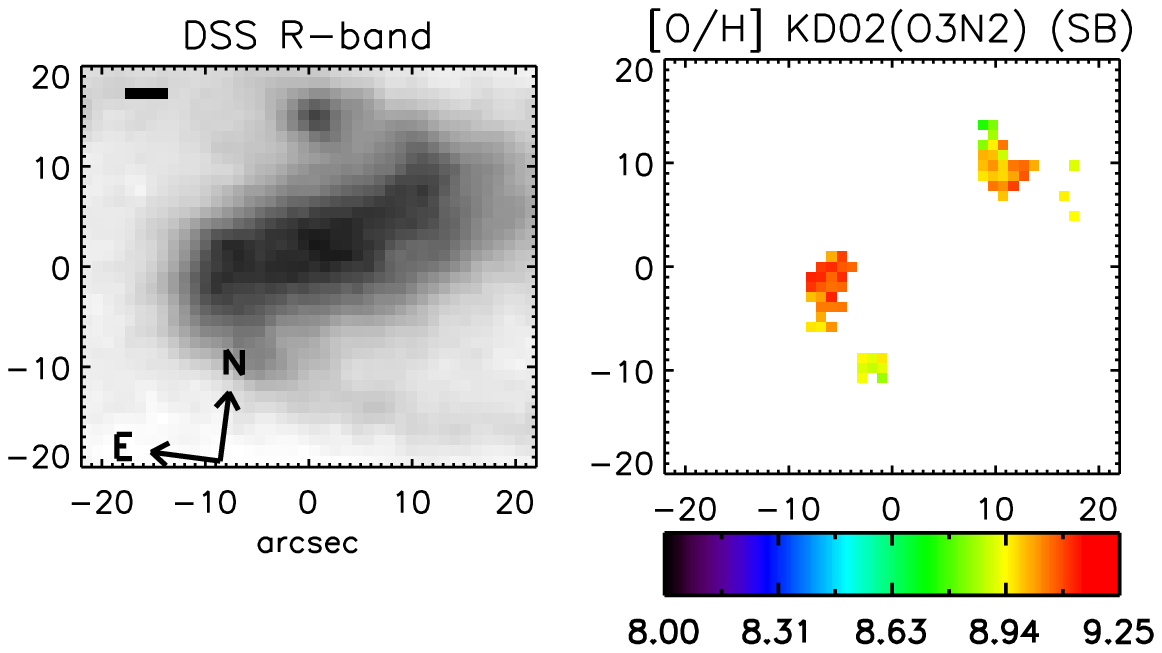}}
{\includegraphics[scale=0.99]{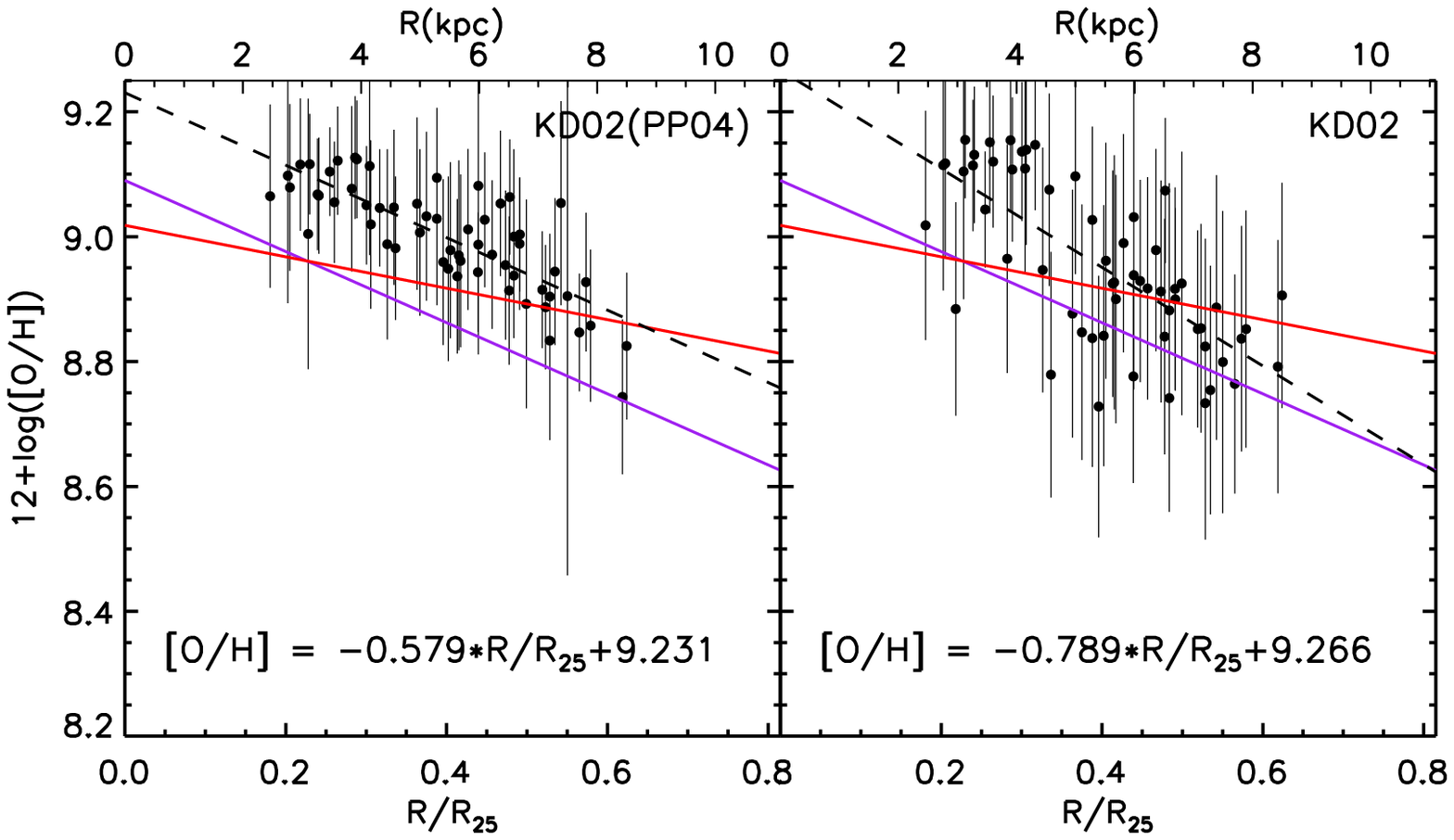}}
\caption{Same as fig A1 for IRAS F18341-5732}

\end{figure*}

\begin{figure*}
\centering
{\includegraphics[scale=0.54]{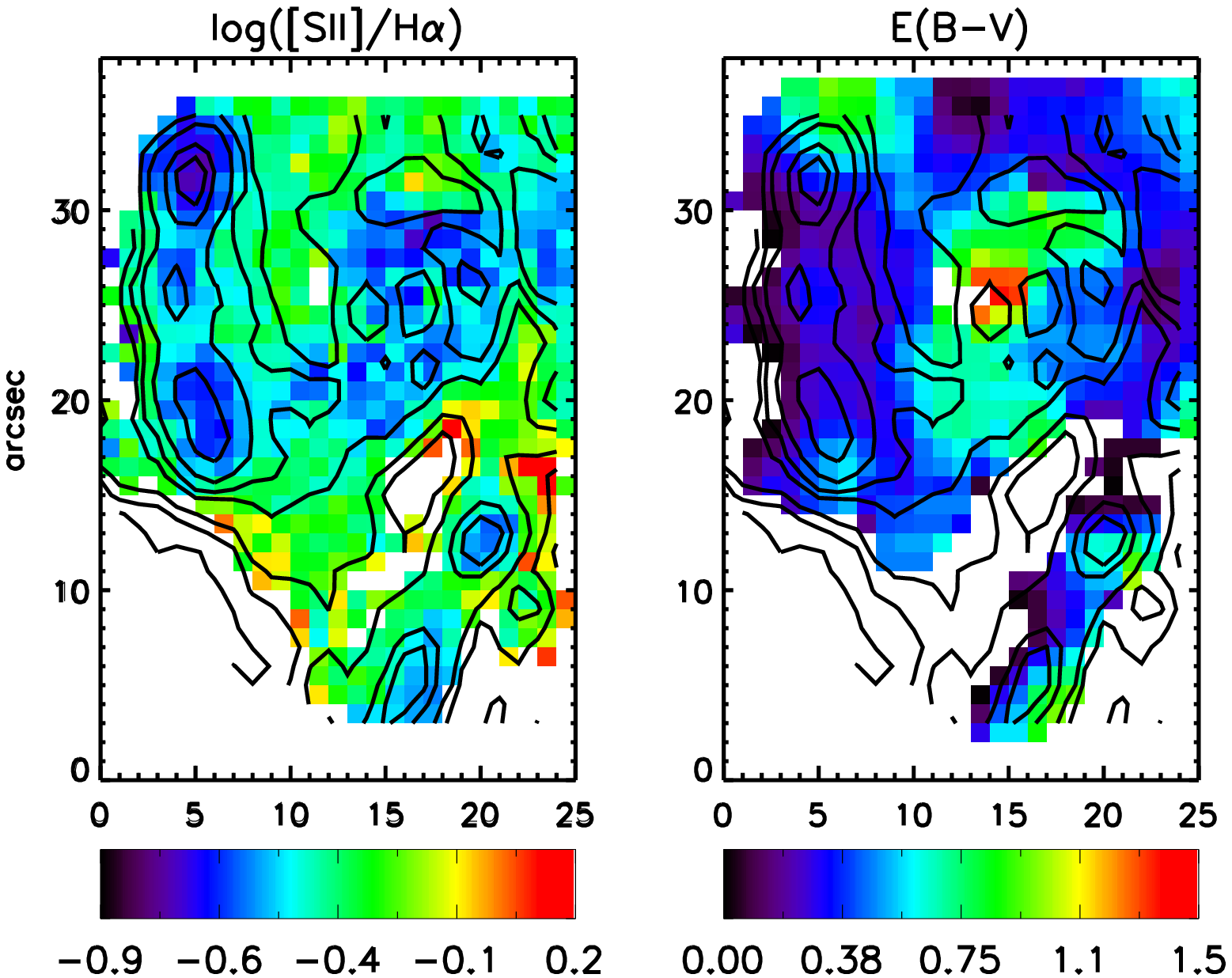}}
{\includegraphics[scale=0.705]{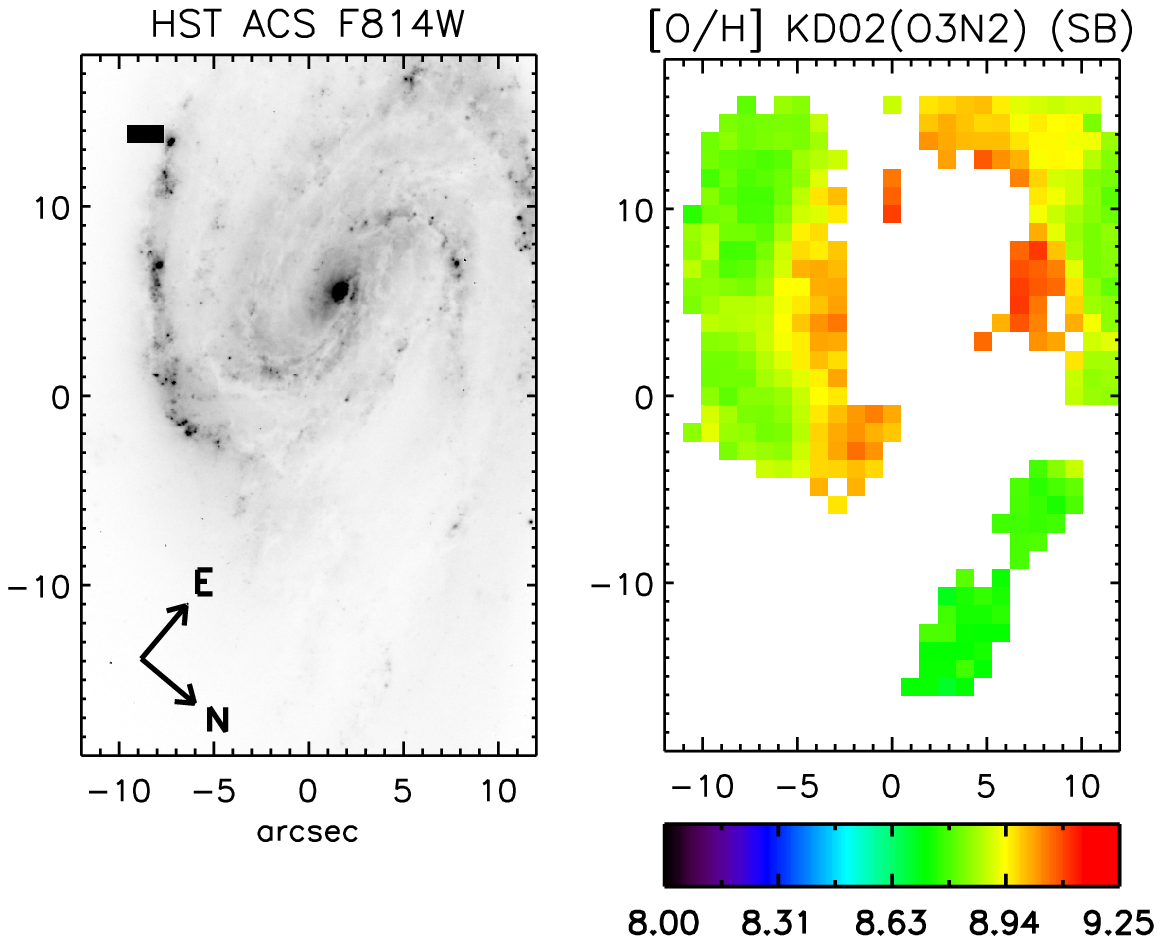}}
{\includegraphics[scale=0.99]{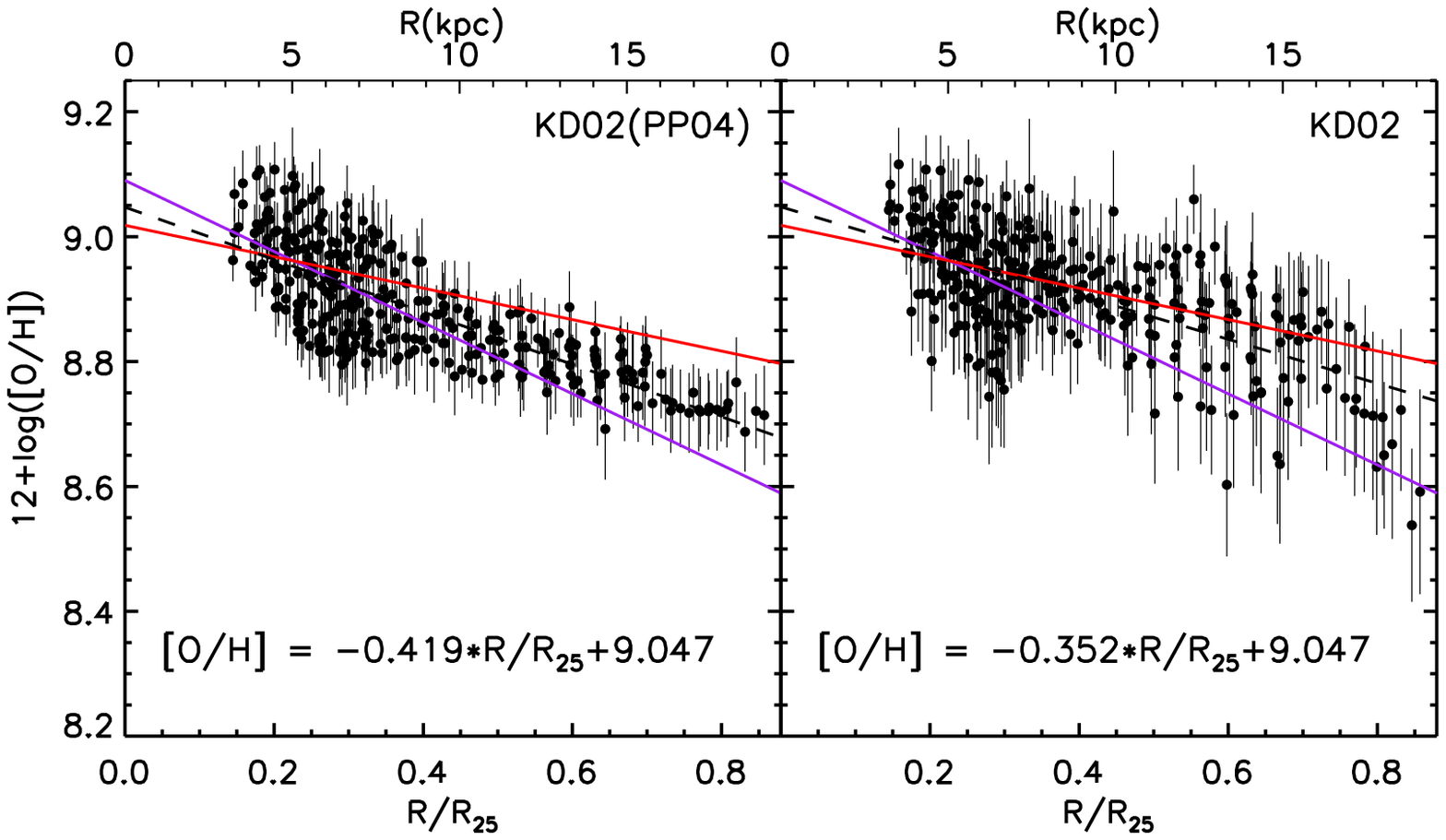}}
\caption{Same as fig A1 for IRAS F13373+0105 W}

\end{figure*}

\begin{figure*}
\centering
{\includegraphics[scale=0.55]{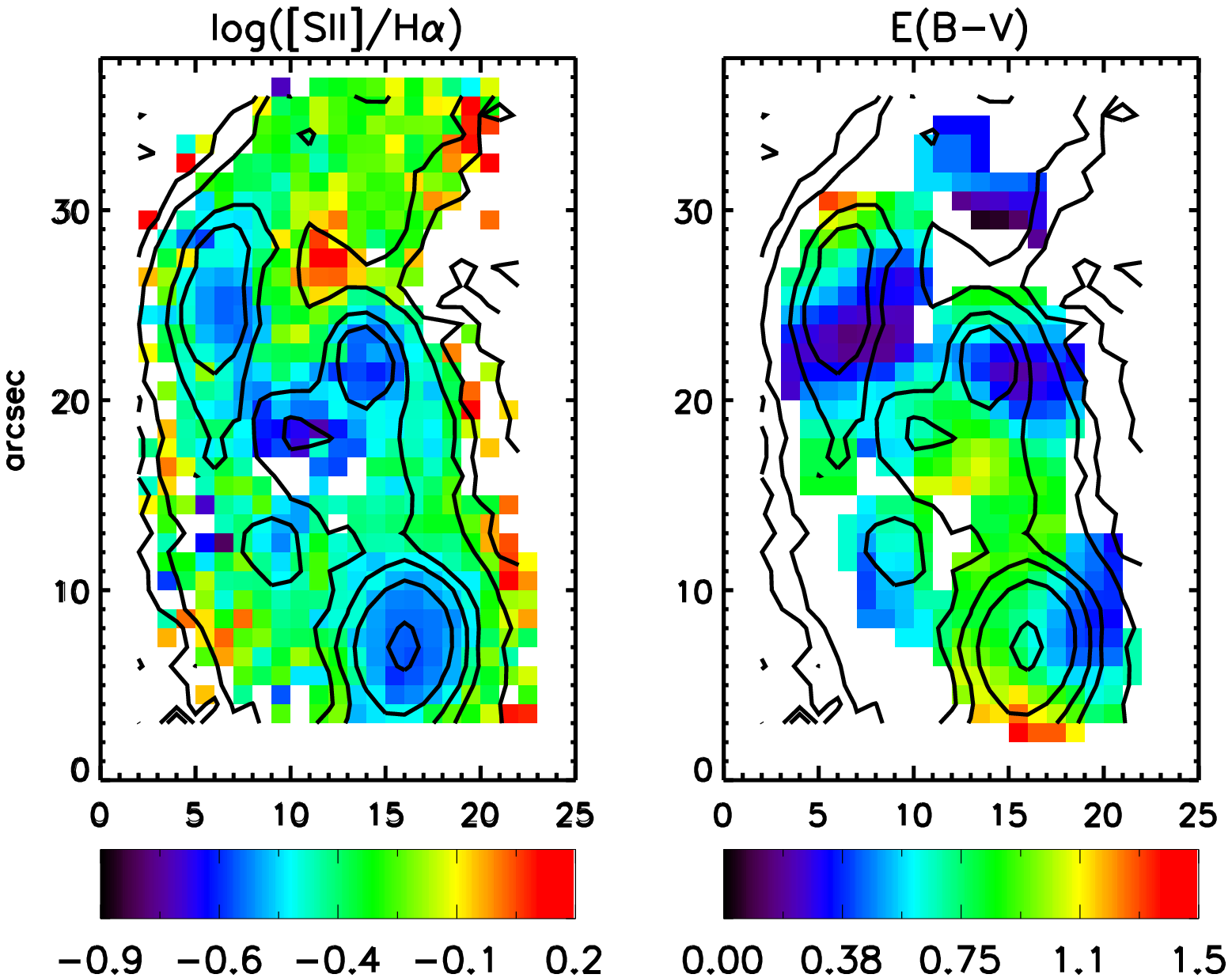}}
{\includegraphics[scale=0.72]{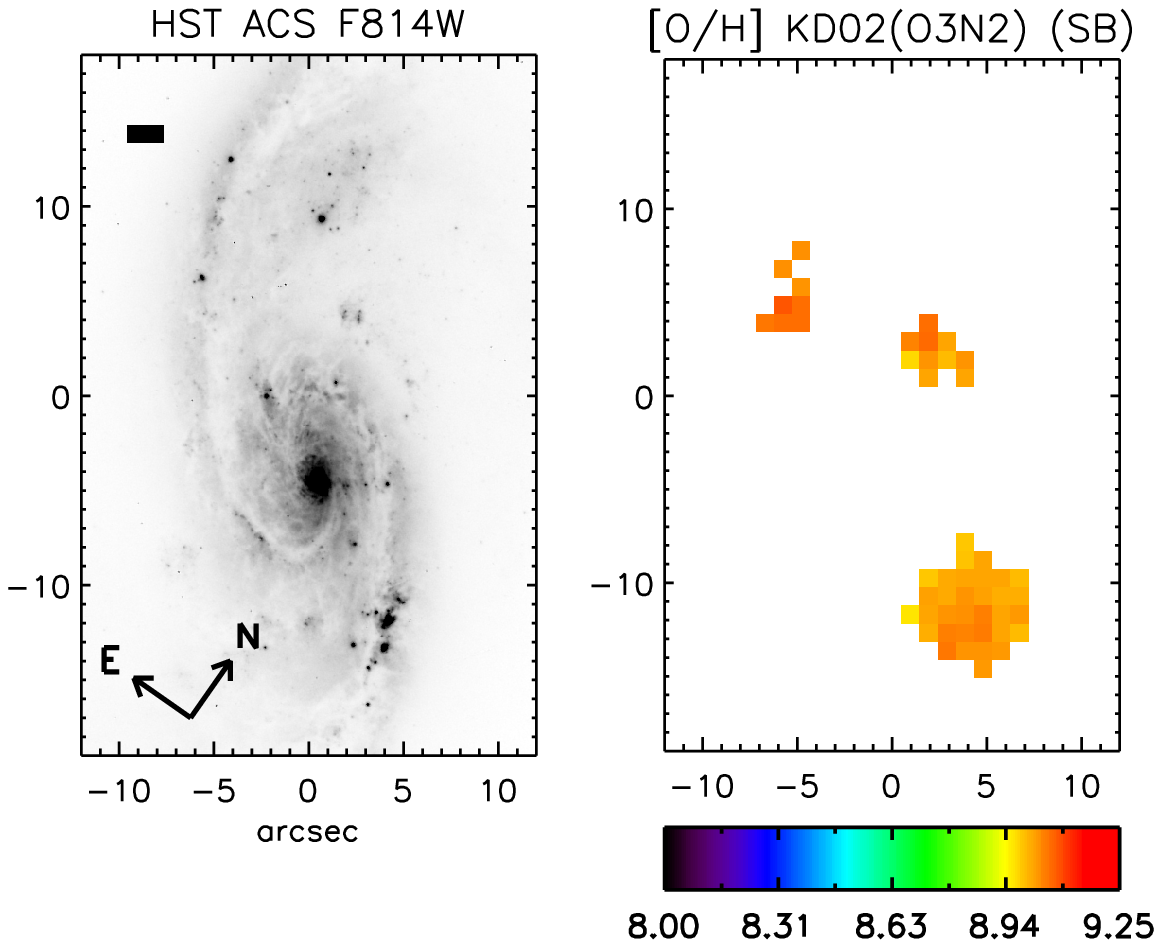}}
{\includegraphics[scale=0.99]{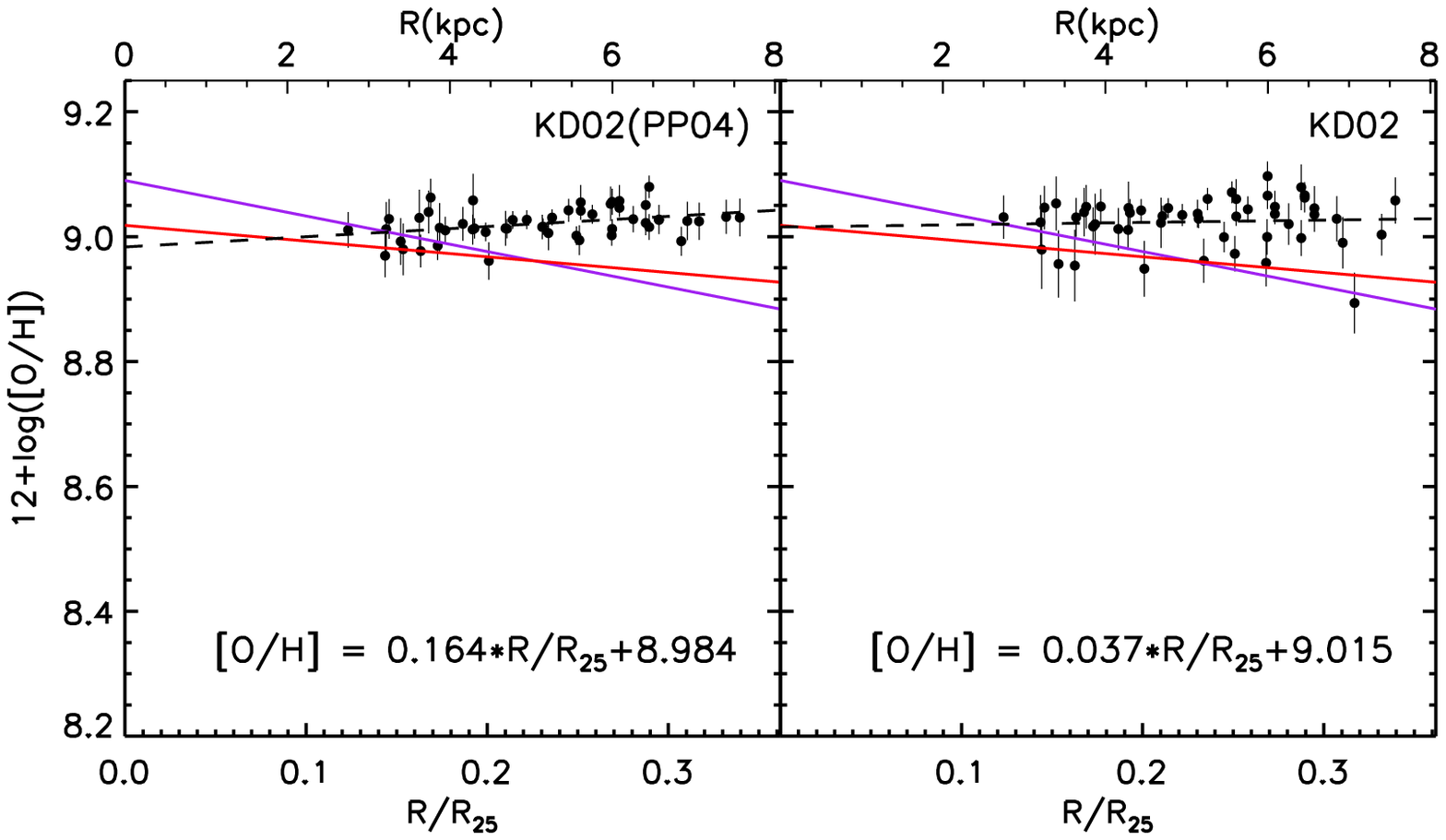}}
\caption{Same as fig A1 for IRAS F13373+0105 E}
\end{figure*}

\begin{figure*}
\centering
{\includegraphics[scale=0.50]{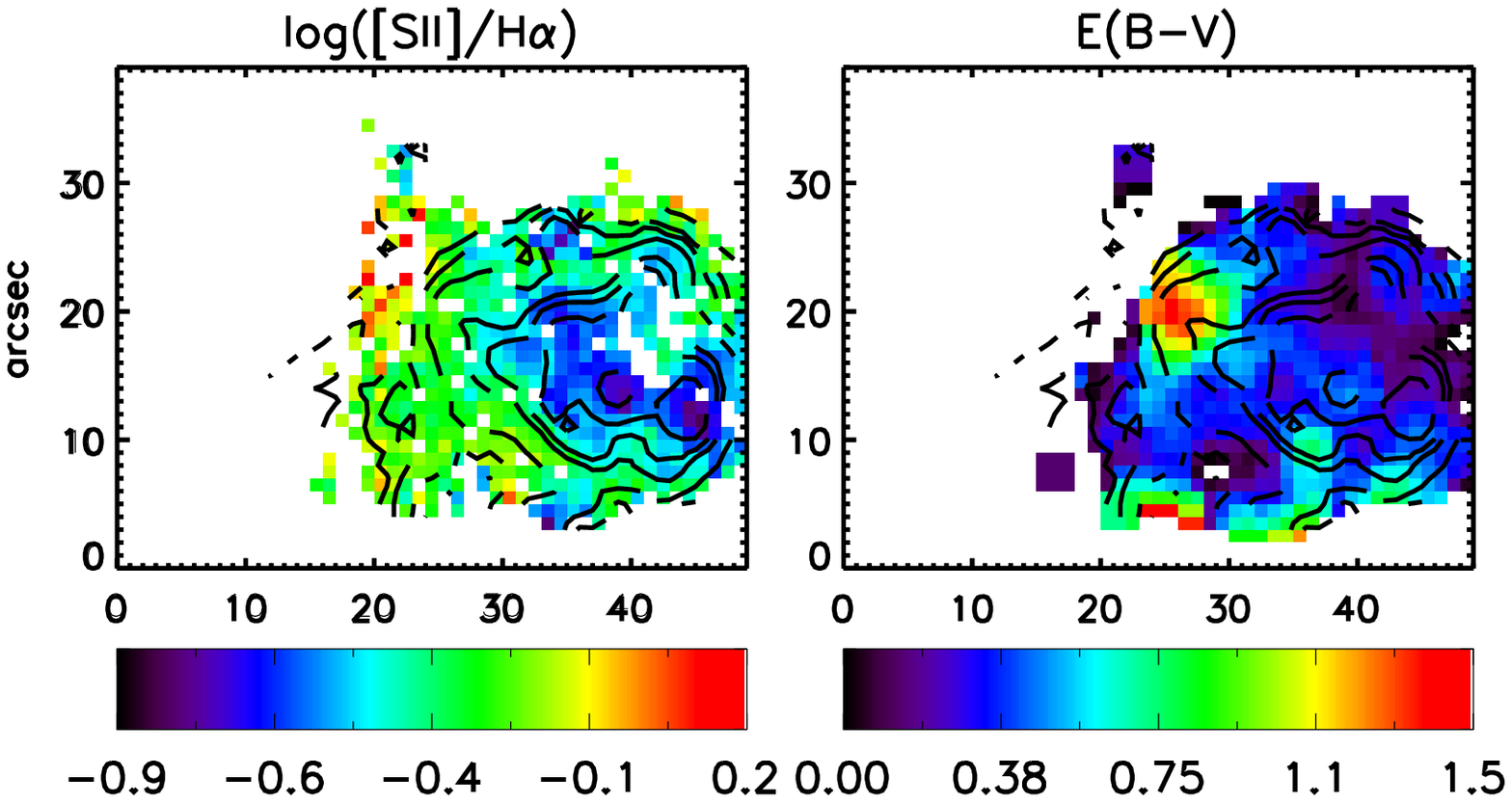}}
{\includegraphics[scale=0.75]{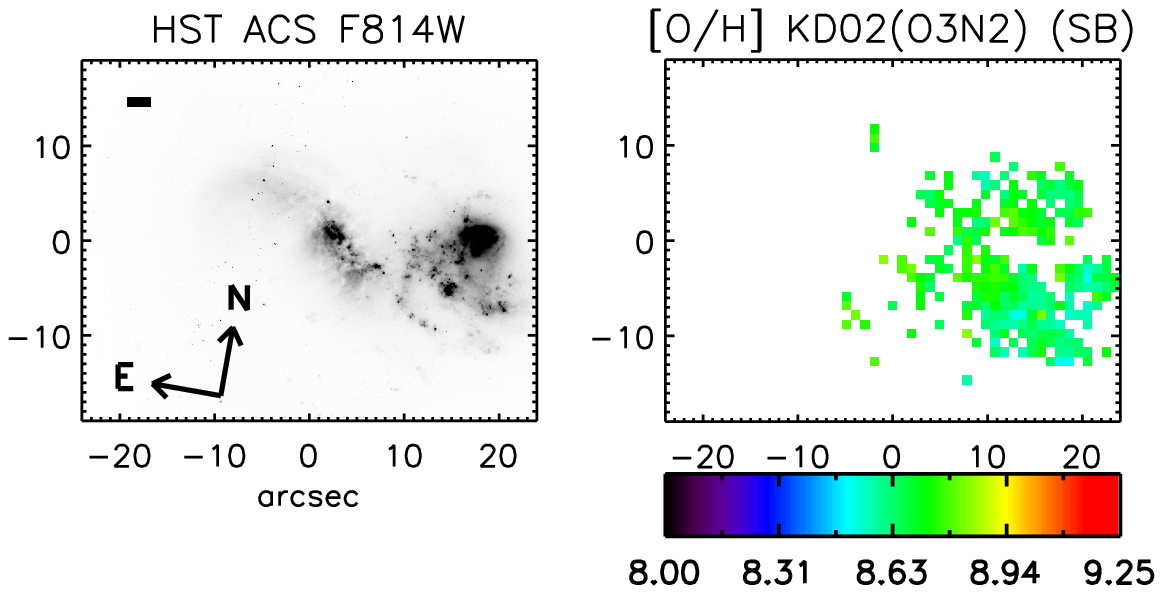}}
{\includegraphics[scale=0.99]{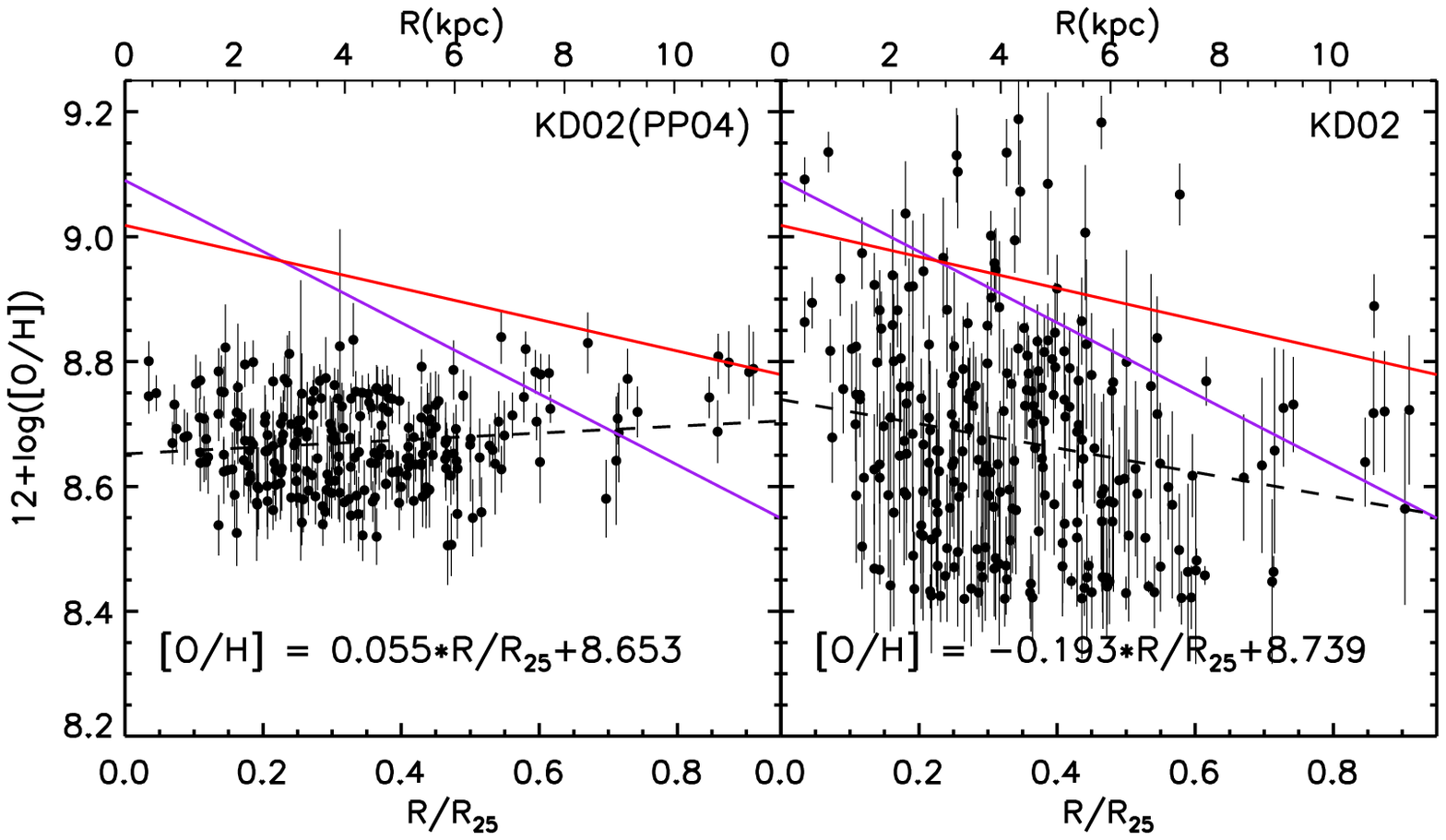}}
\caption{Same as fig A1 for IRAS F01053-1746}
\end{figure*}

\begin{figure*}
\centering
{\includegraphics[scale=0.515]{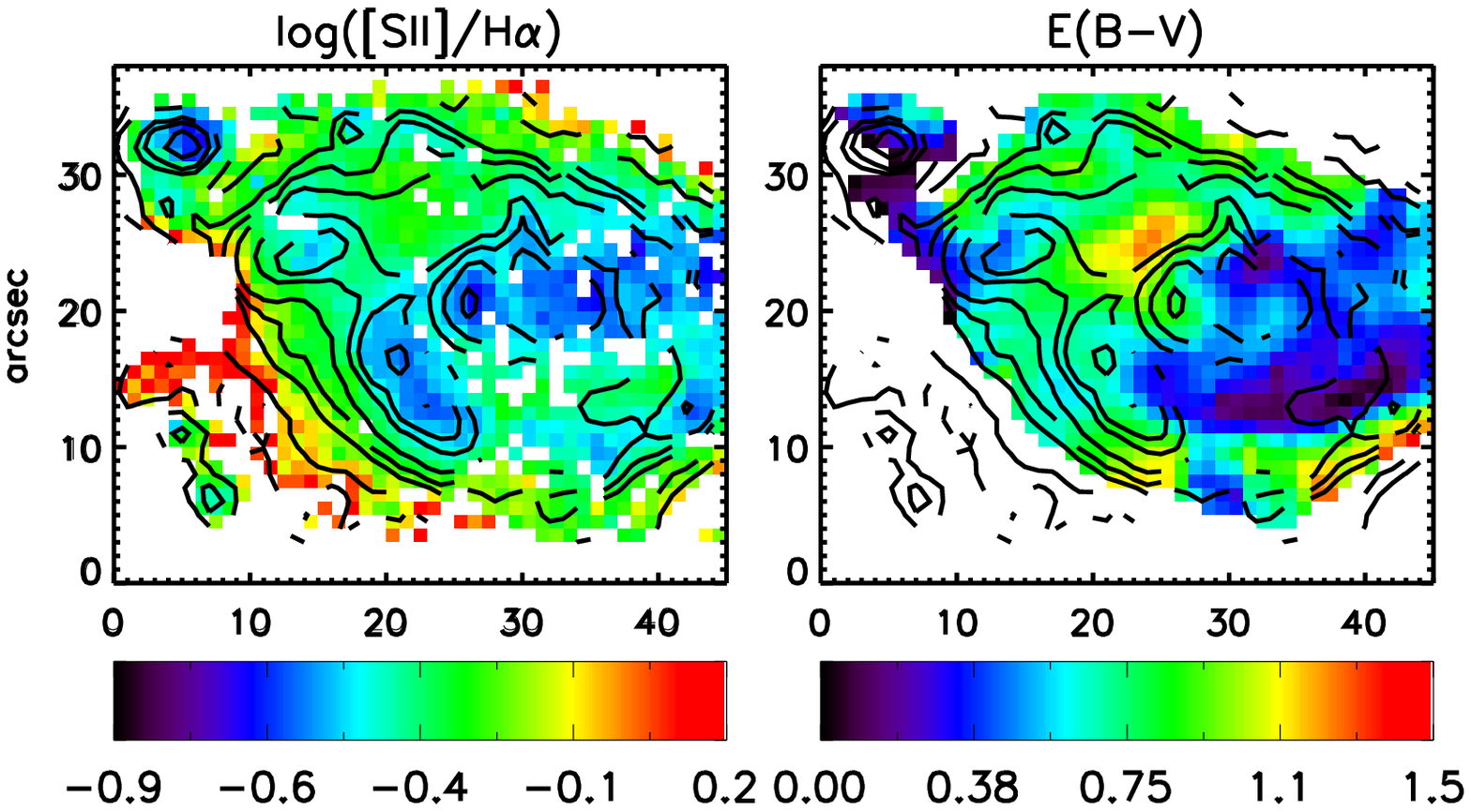}}
{\includegraphics[scale=0.77]{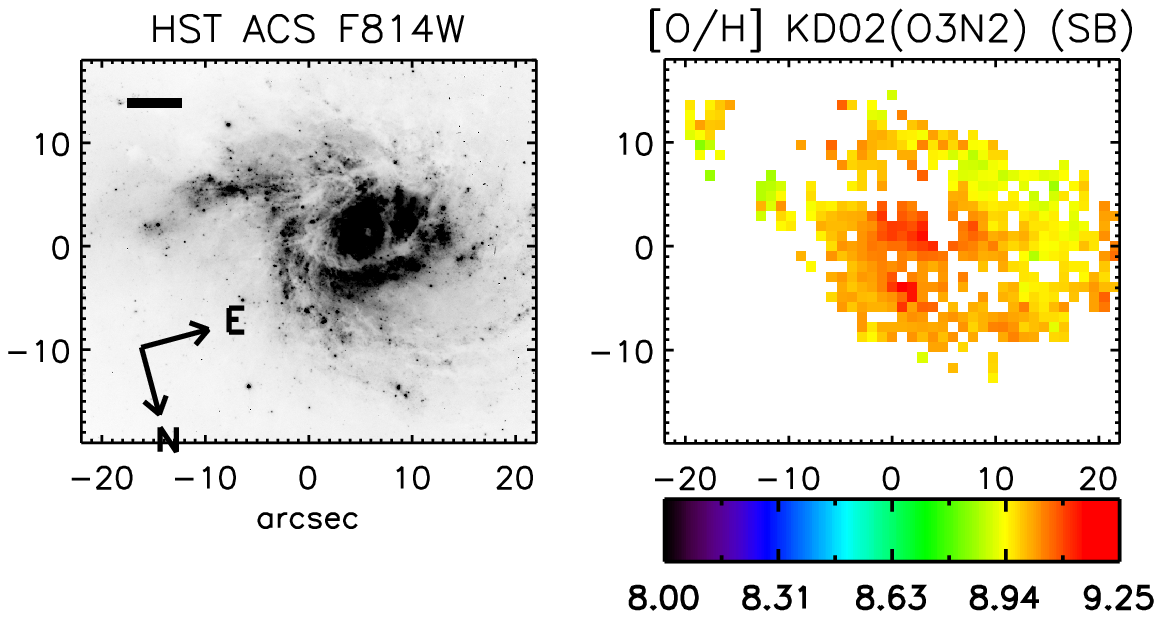}}
{\includegraphics[scale=0.99]{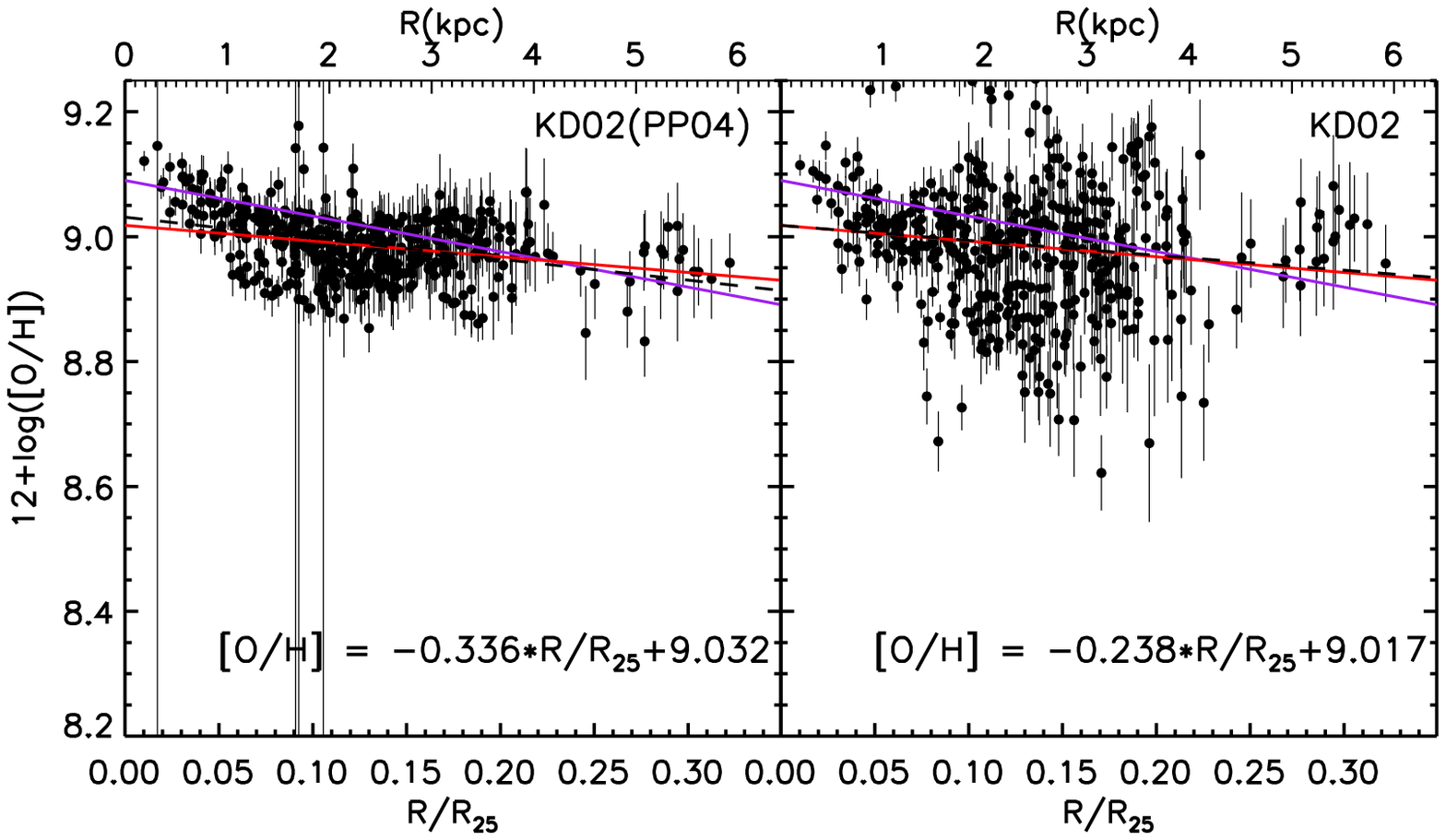}}
\caption{Same as fig A1 for IRAS F10257-4339}
\end{figure*}

\begin{figure*}
\centering
{\includegraphics[scale=0.52]{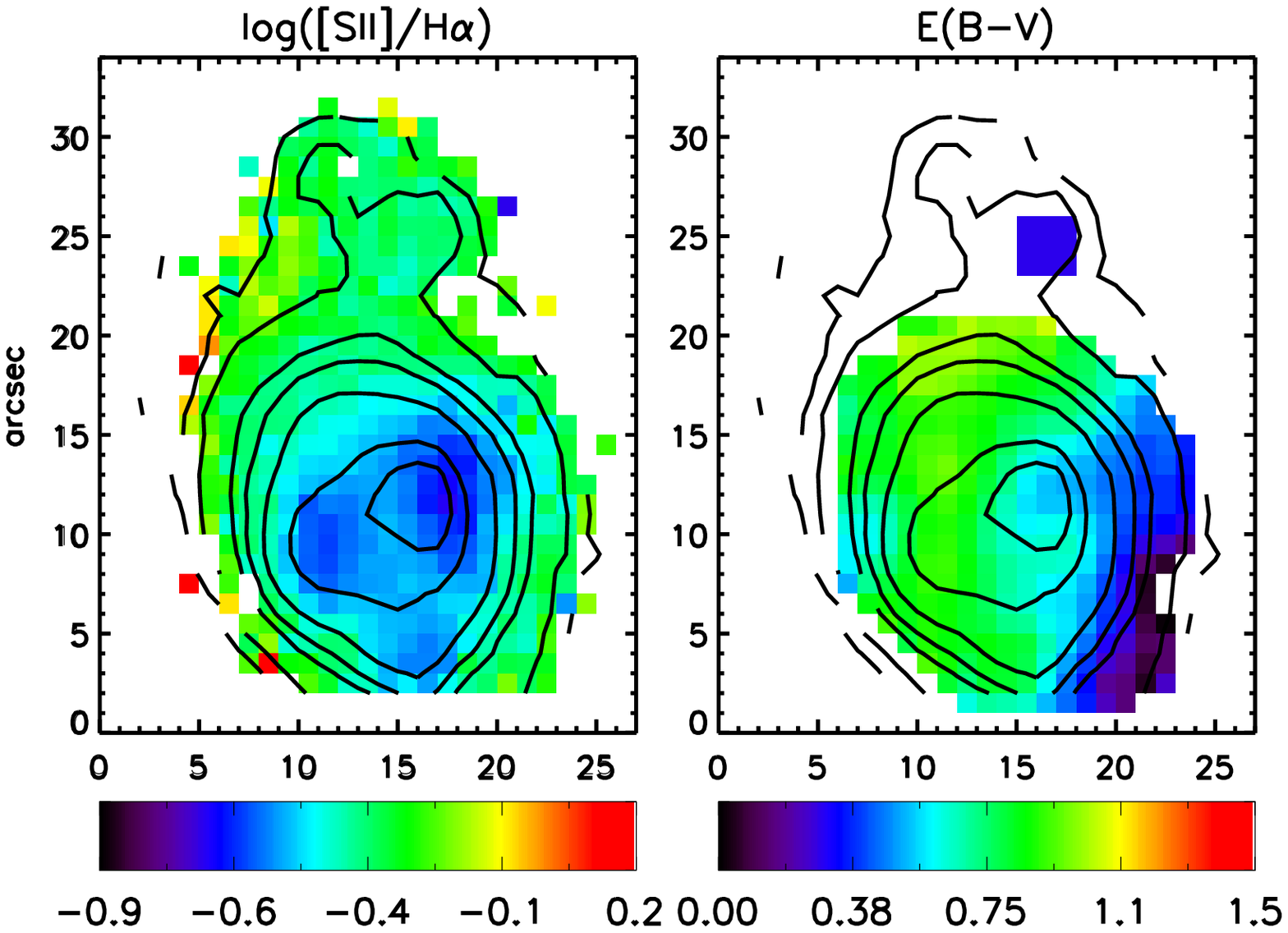}}
{\includegraphics[scale=0.773]{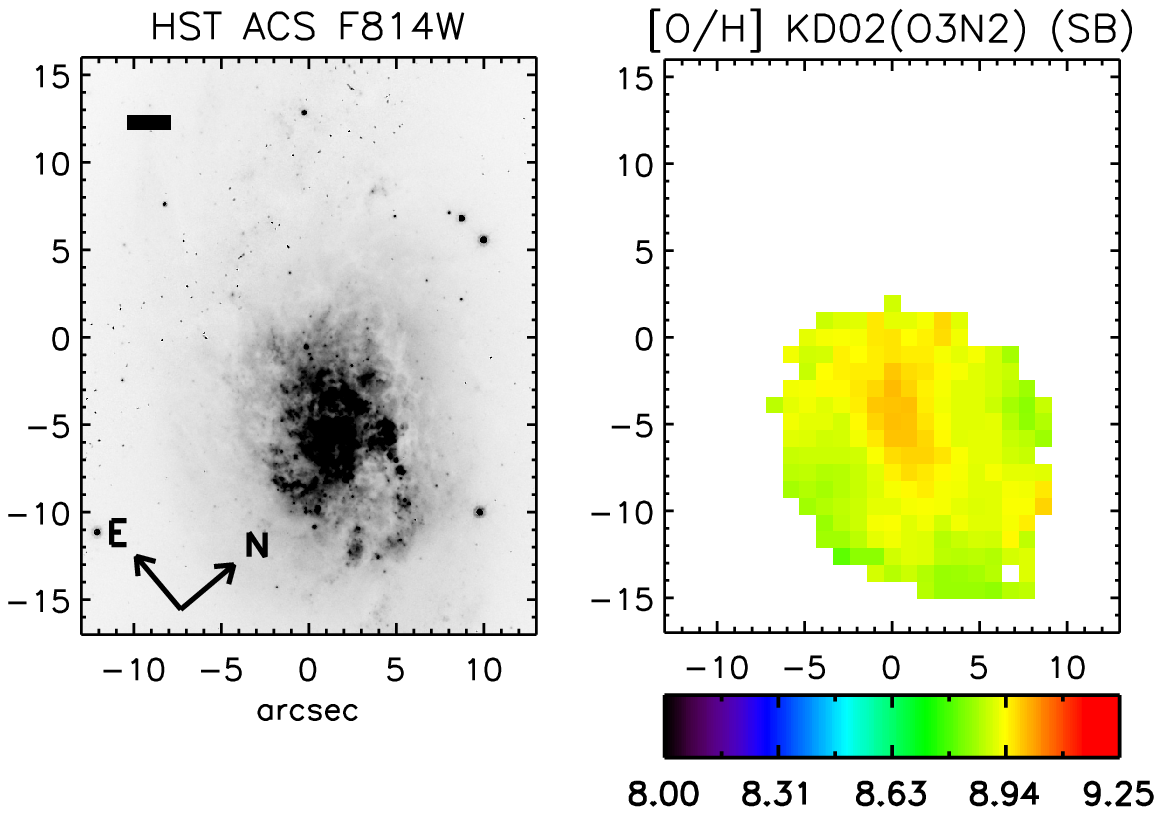}}
{\includegraphics[scale=0.99]{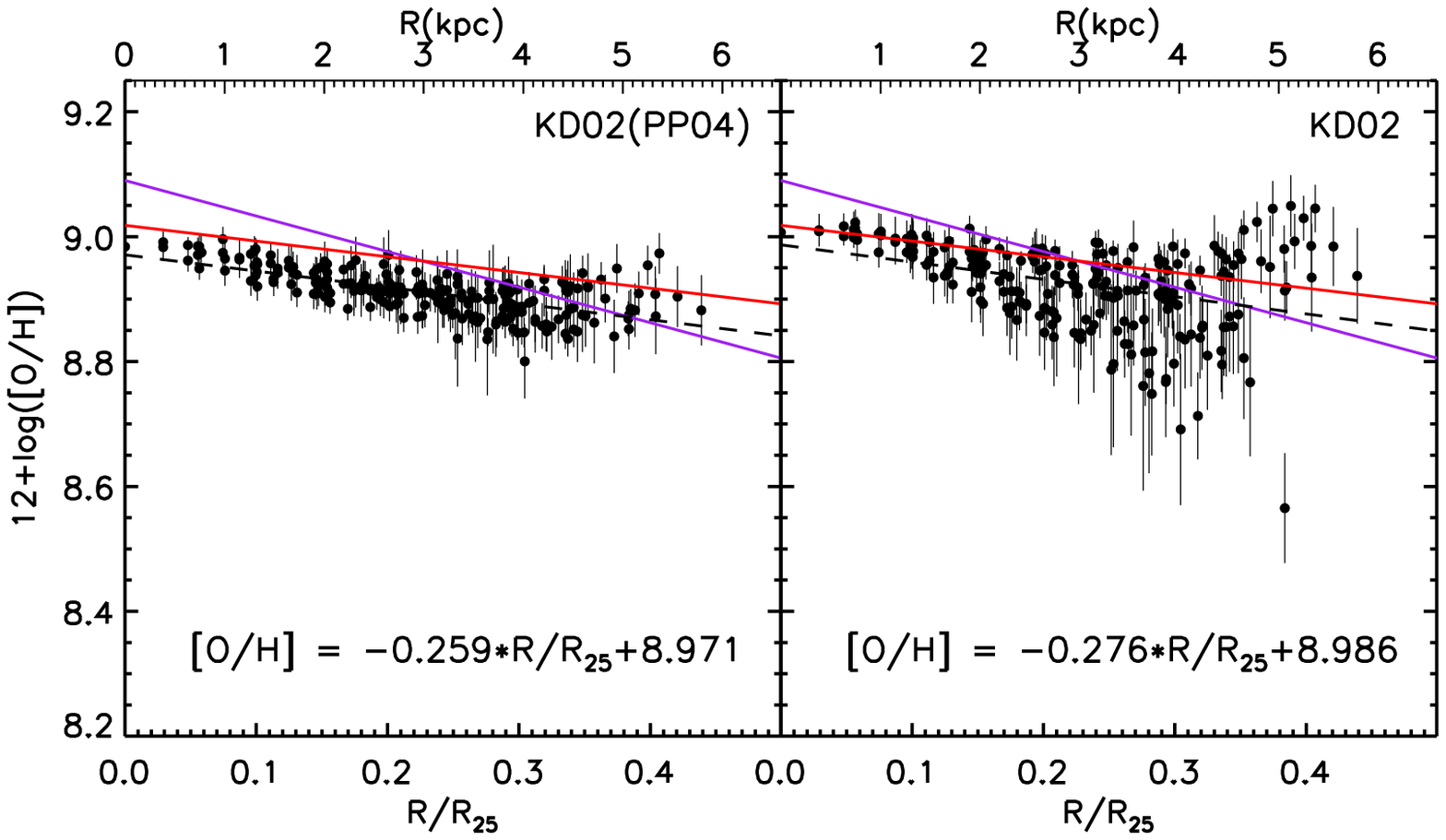}}
\caption{Same as fig A1 for IRAS F18093-5744 N}
\end{figure*}

\begin{figure*}
\centering
{\includegraphics[scale=0.54]{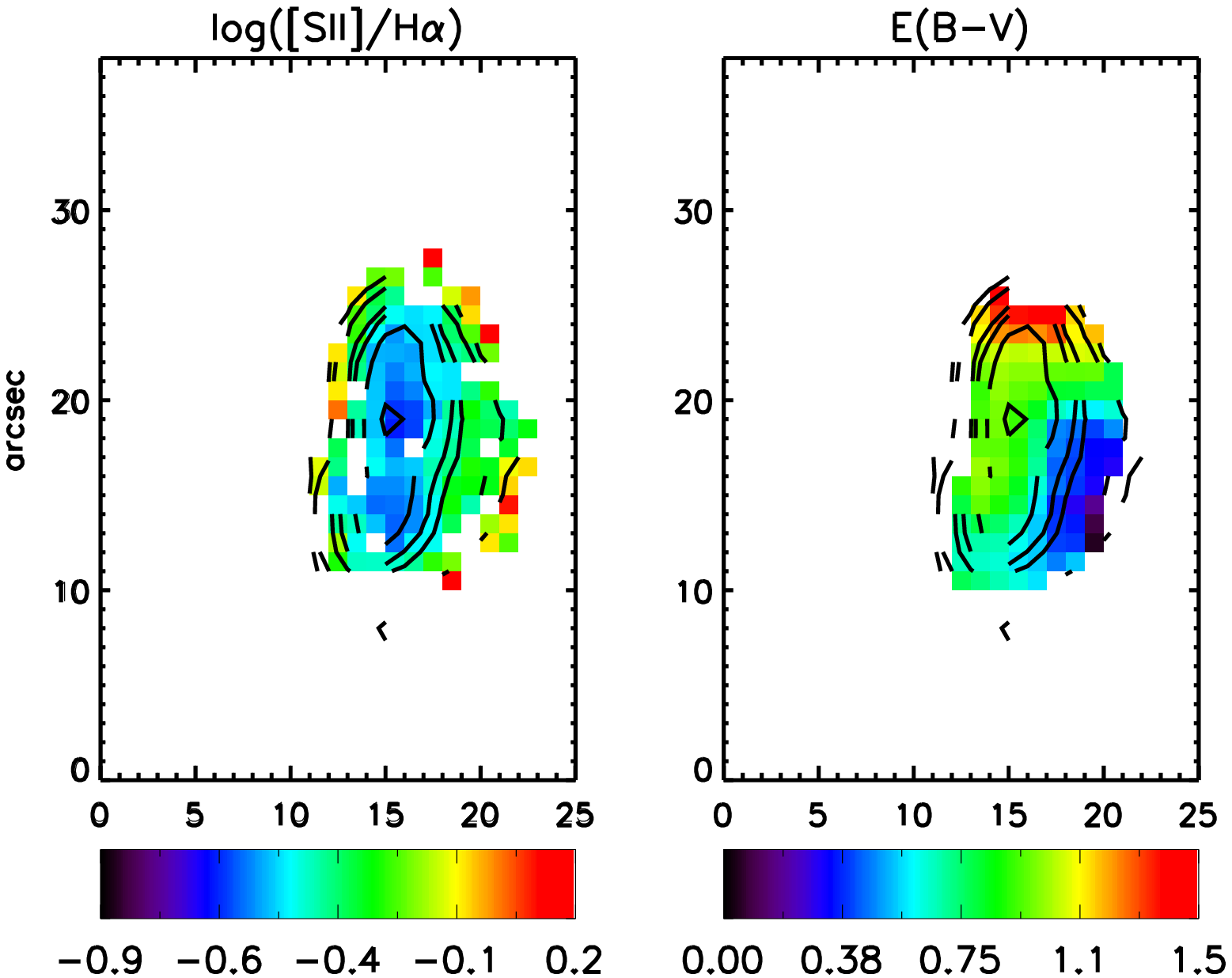}}
{\includegraphics[scale=0.705]{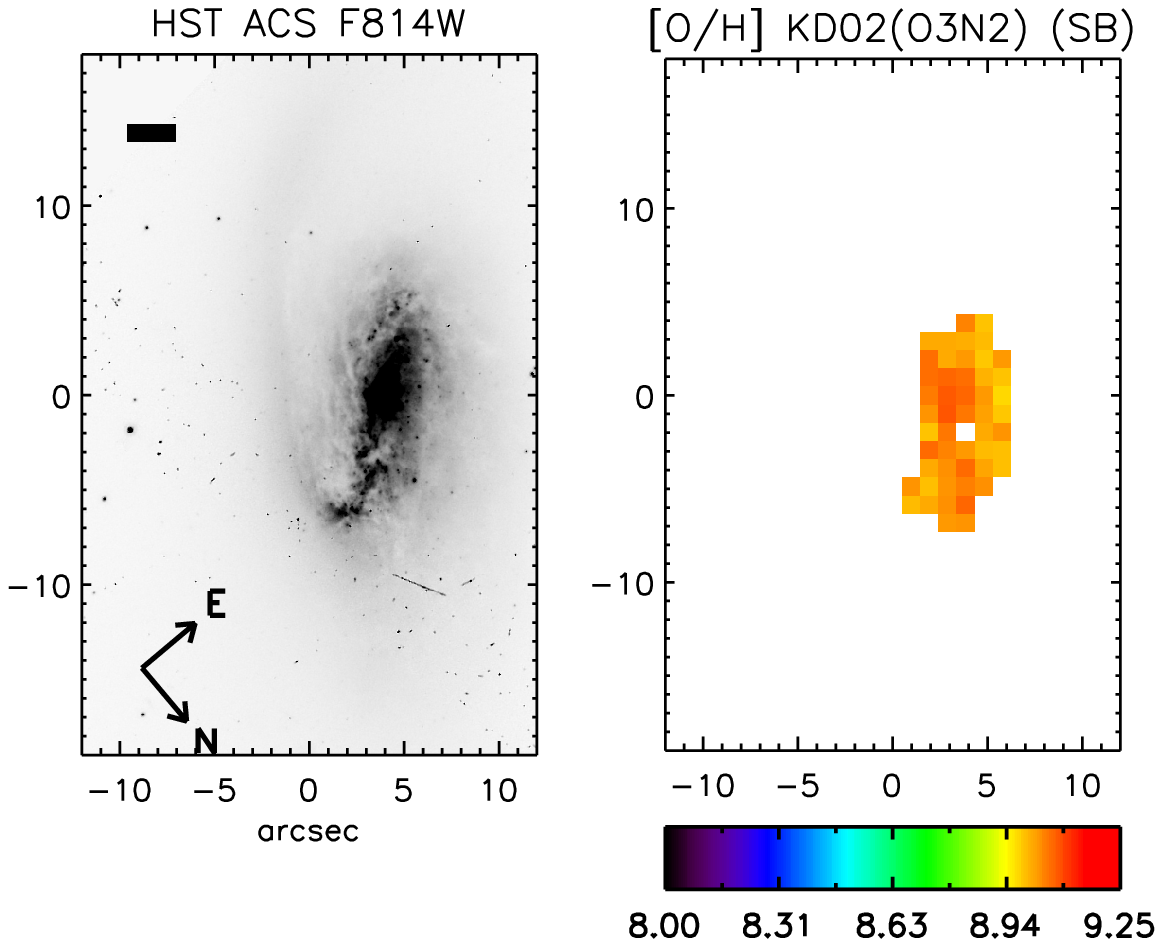}}
{\includegraphics[scale=0.99]{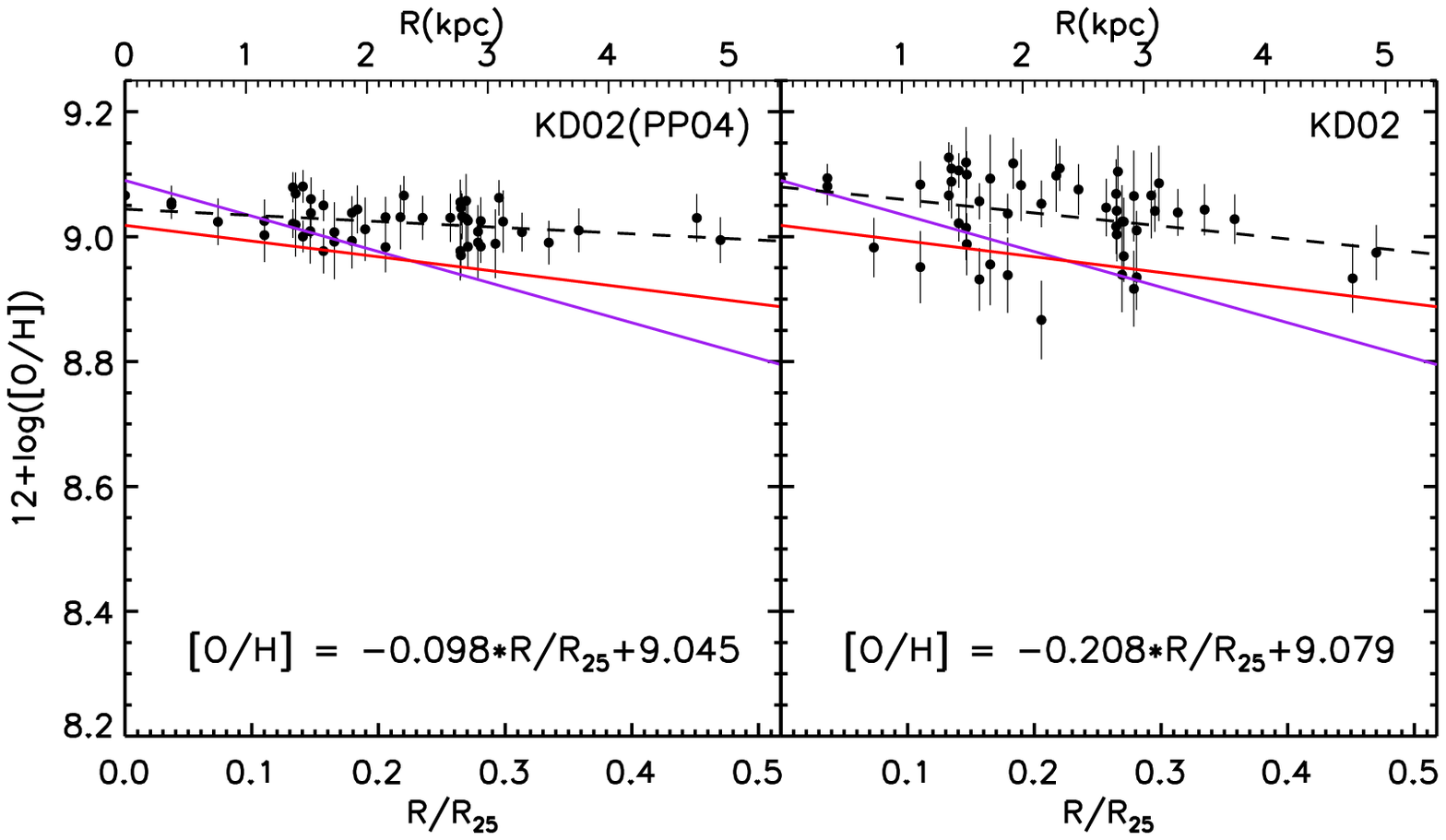}}
\caption{Same as fig A1 for IRAS F18093-5744 S}
\end{figure*}

\begin{figure*}
\centering
{\includegraphics[scale=0.50]{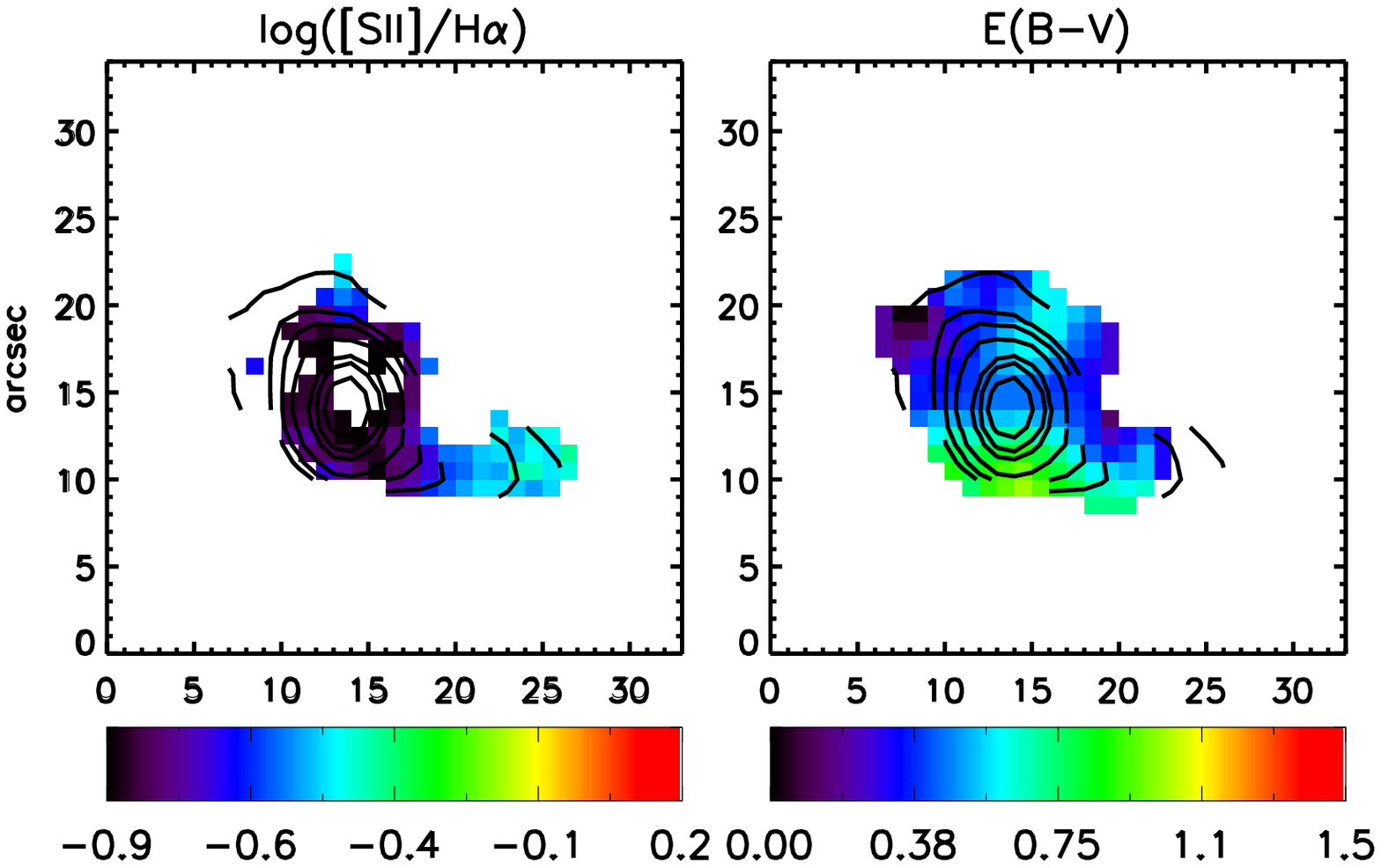}}
{\includegraphics[scale=0.75]{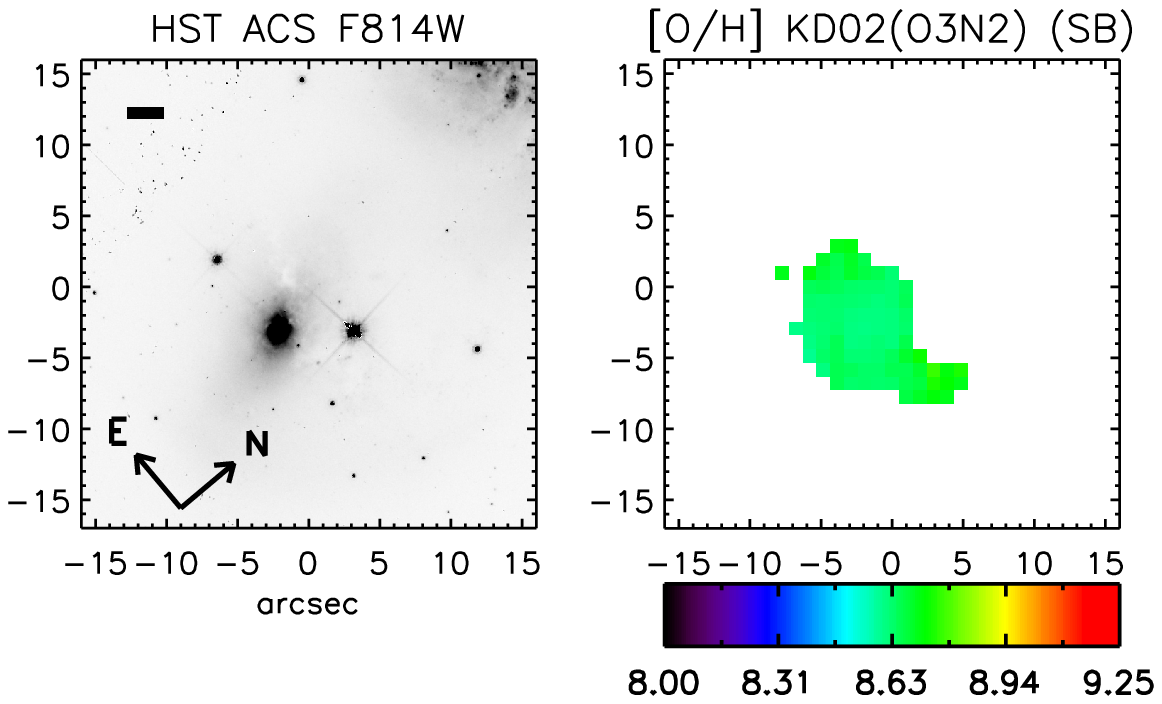}}
{\includegraphics[scale=0.99]{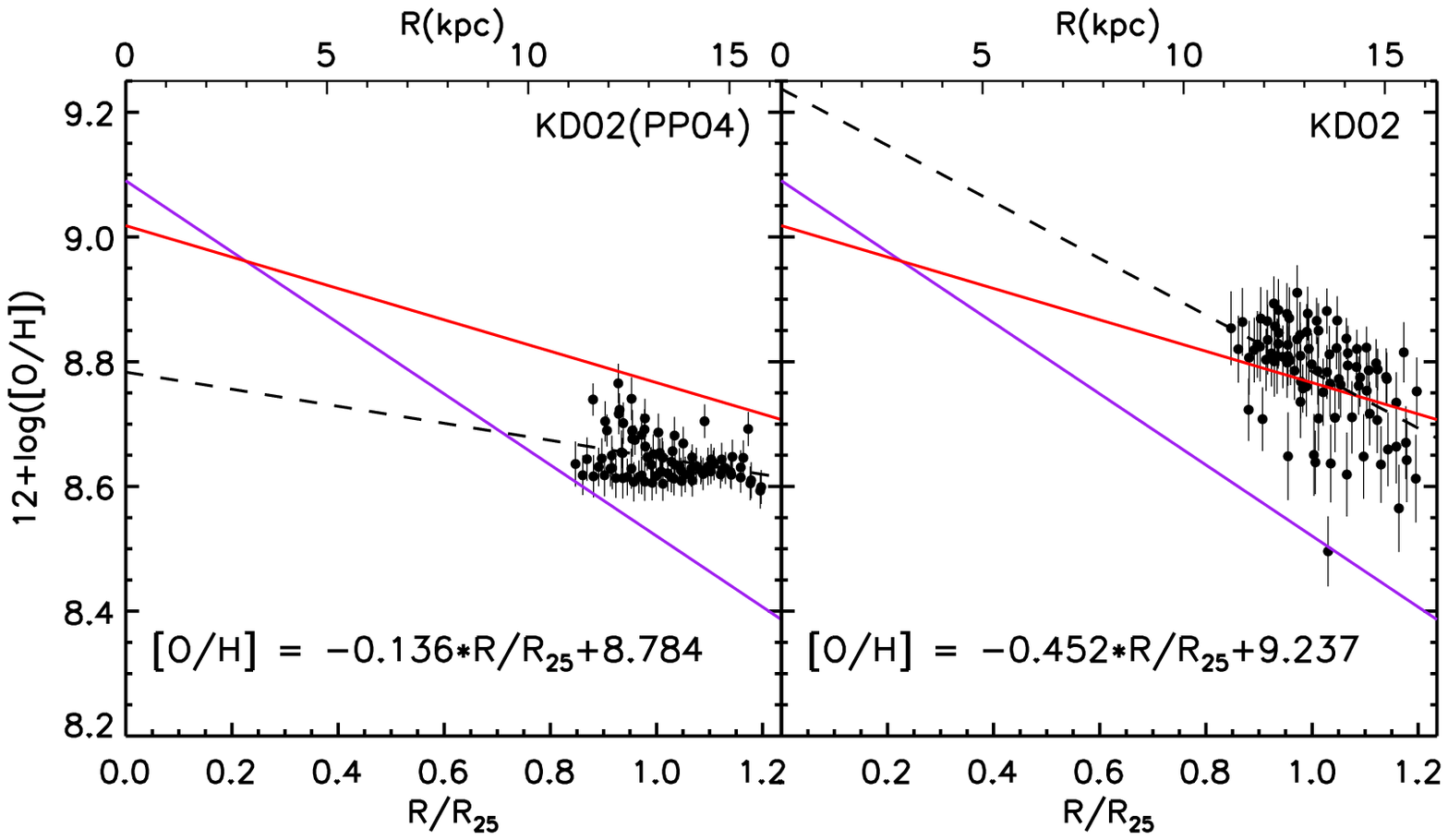}}
\caption{Same as fig A1 for IRAS F18093-5744 C. As discussed in the text, this appears to be a mix of IRAS F18093-5744 N and a WR galaxy, radii are plotted from the center of IRASF 18093-5744 and the derived values are not included in our final analysis.}
\end{figure*}

\begin{figure*}
\centering
{\includegraphics[scale=0.54]{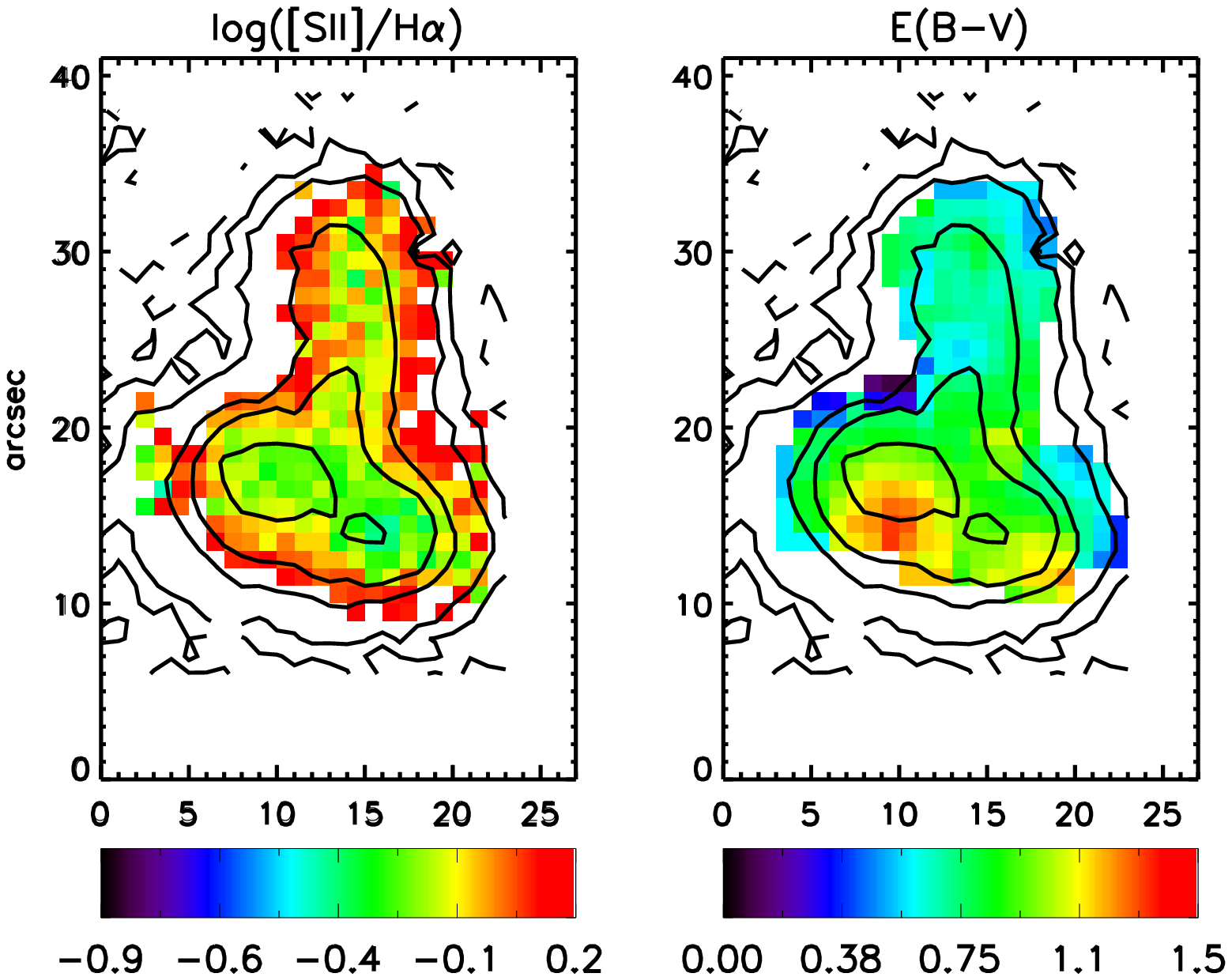}}
{\includegraphics[scale=0.71]{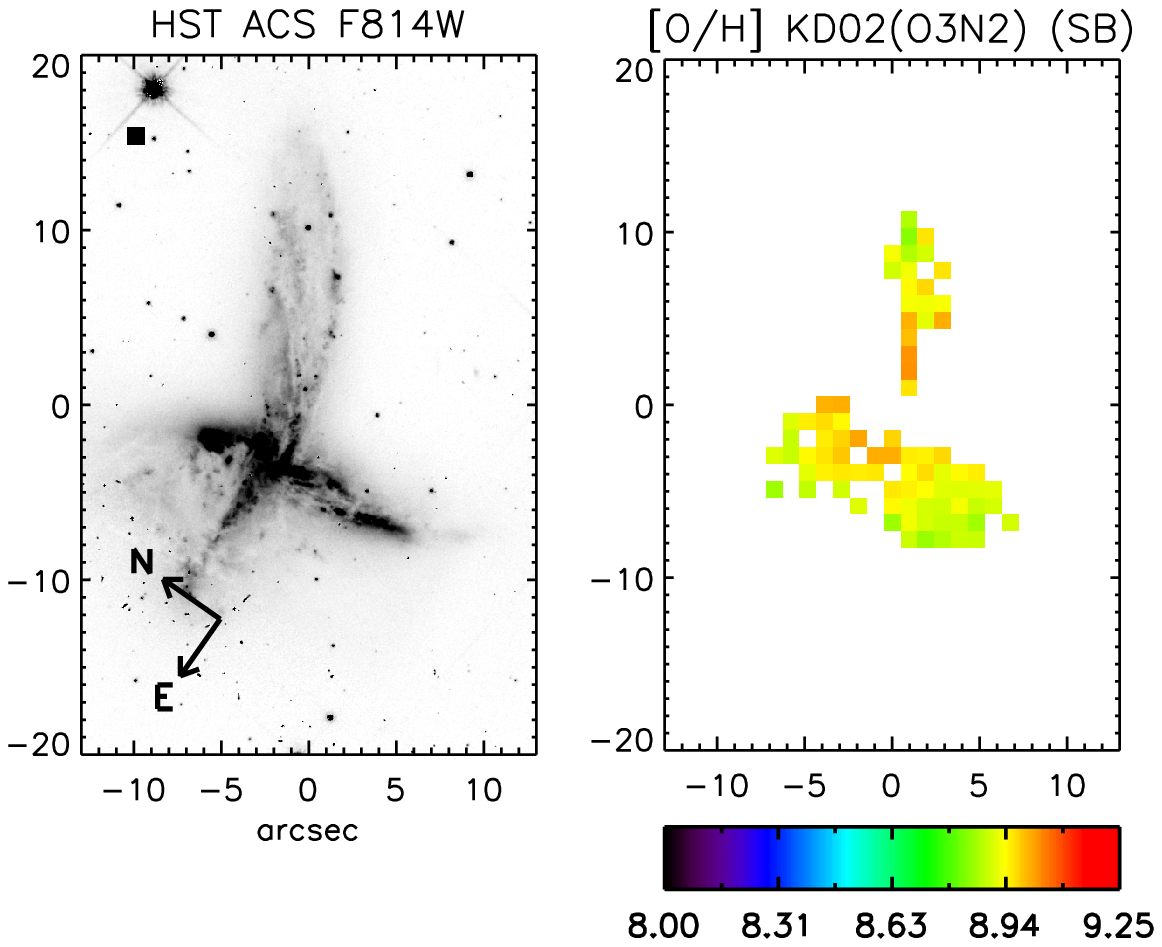}}
{\includegraphics[scale=0.99]{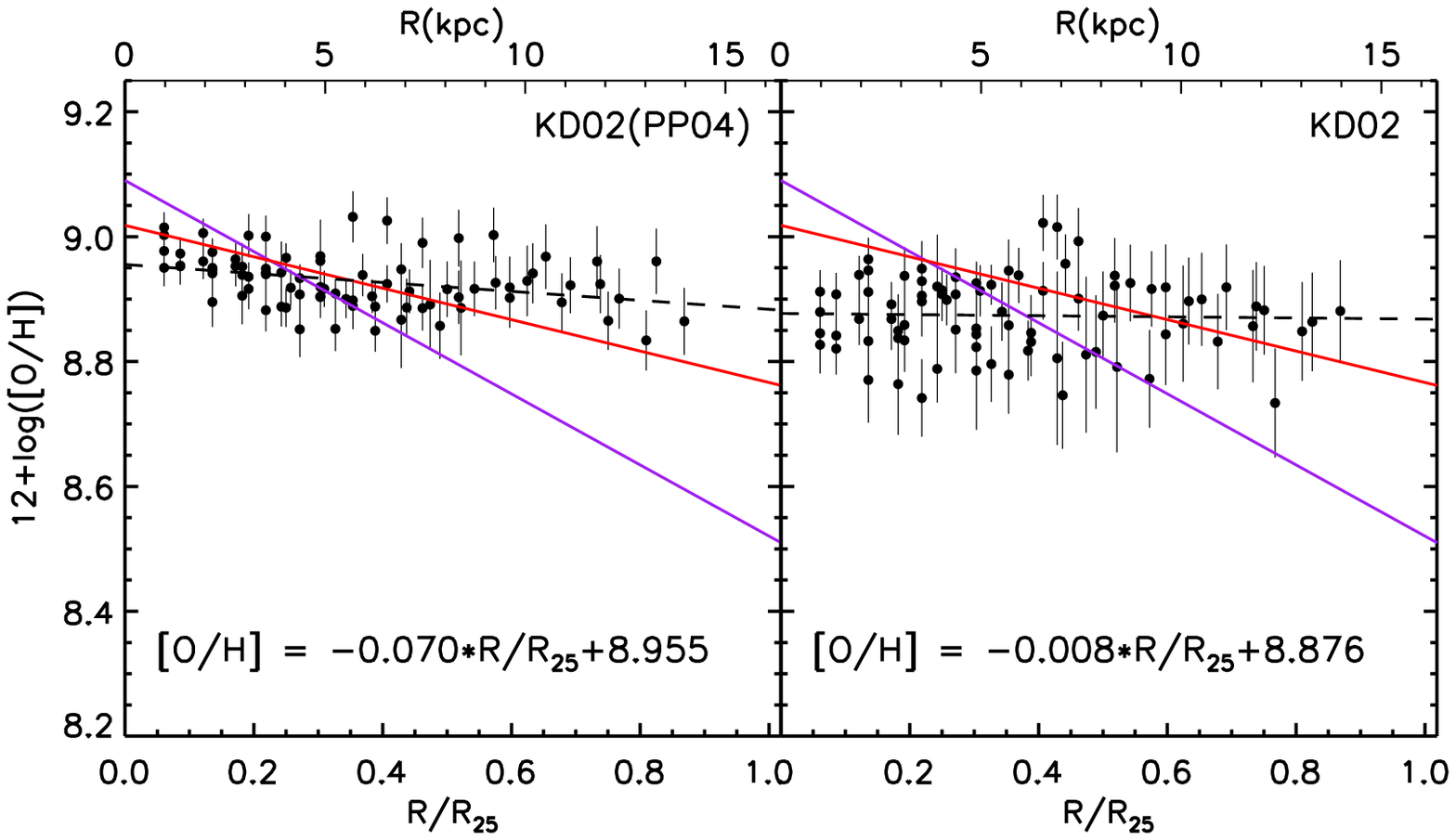}}
\caption{Same as fig A1 for IRAS F19115-2124}
\end{figure*}

\begin{figure*}
\centering
{\includegraphics[scale=0.545]{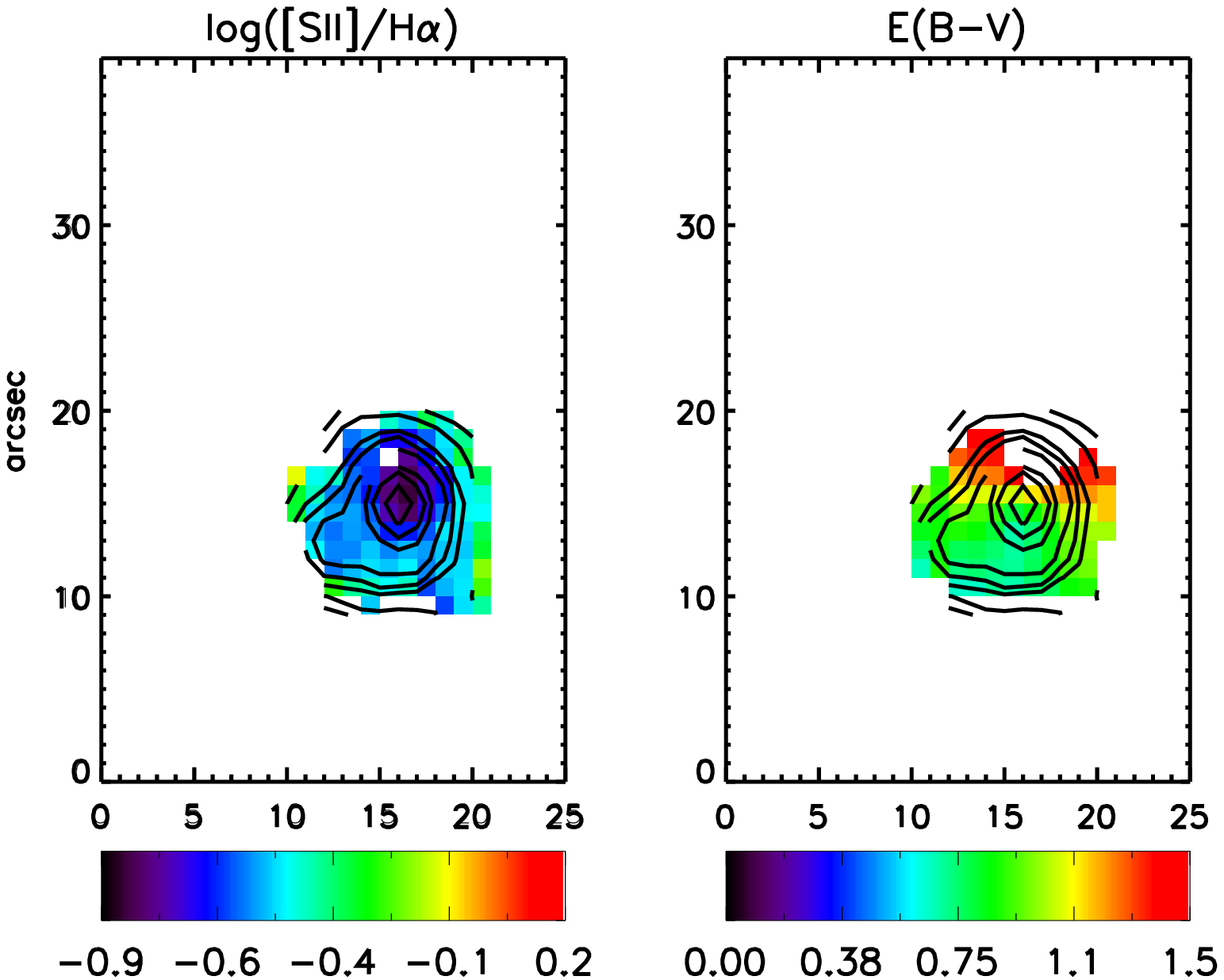}}
{\includegraphics[scale=0.692]{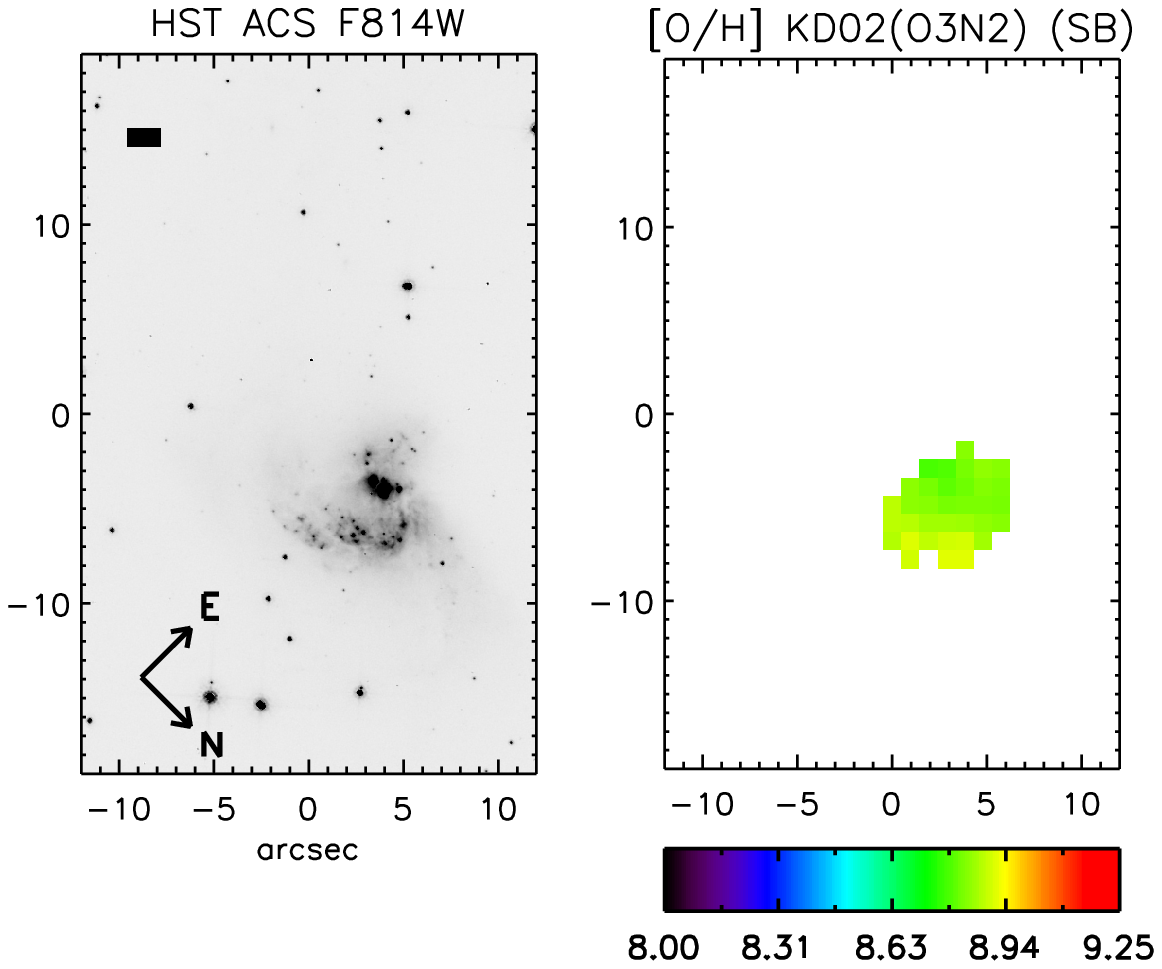}}
{\includegraphics[scale=0.99]{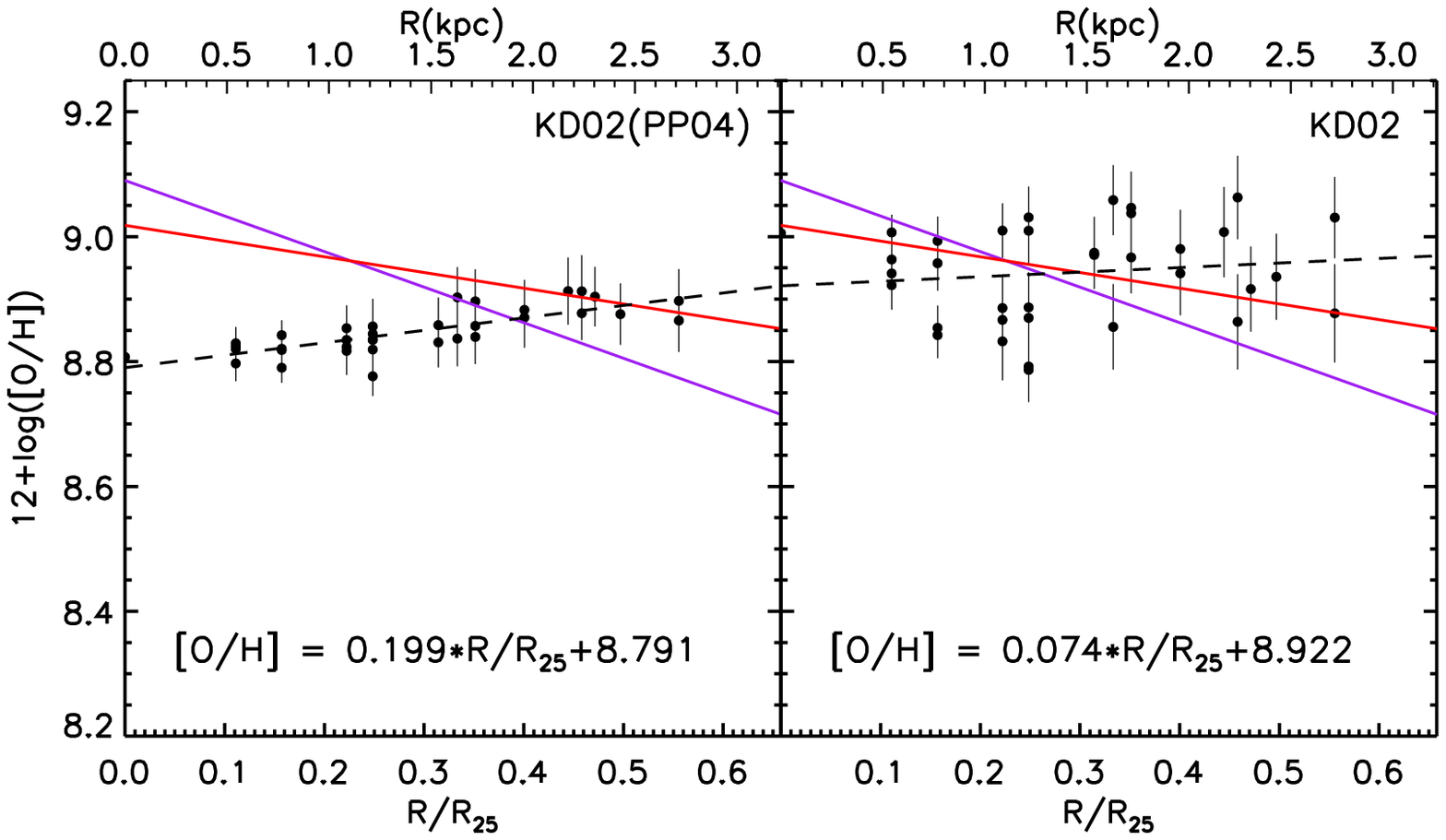}}
\caption{Same as fig A1 for IRAS 08355-4944}
\end{figure*}

\begin{figure*}
\centering
{\includegraphics[scale=0.56]{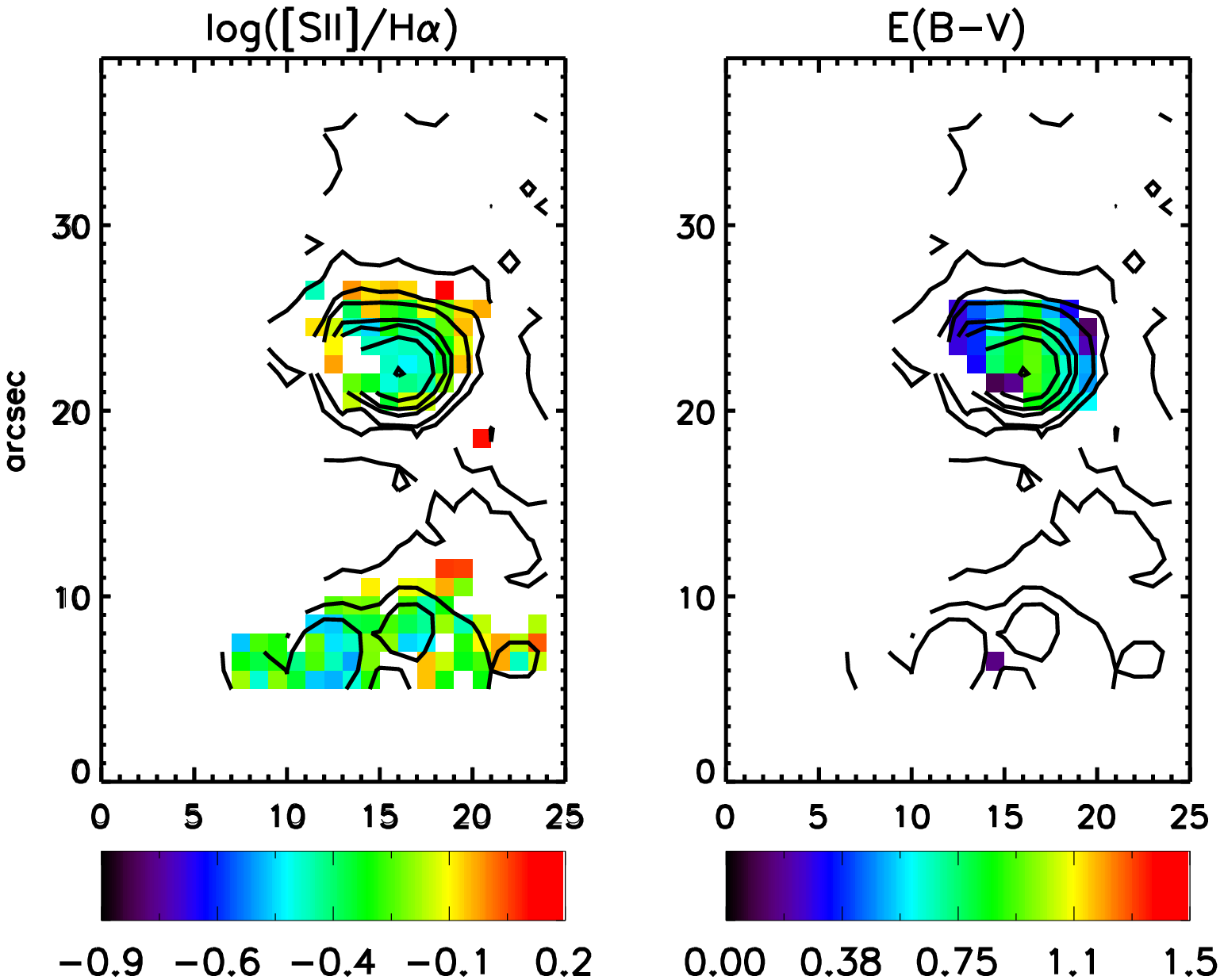}}
{\includegraphics[scale=0.71]{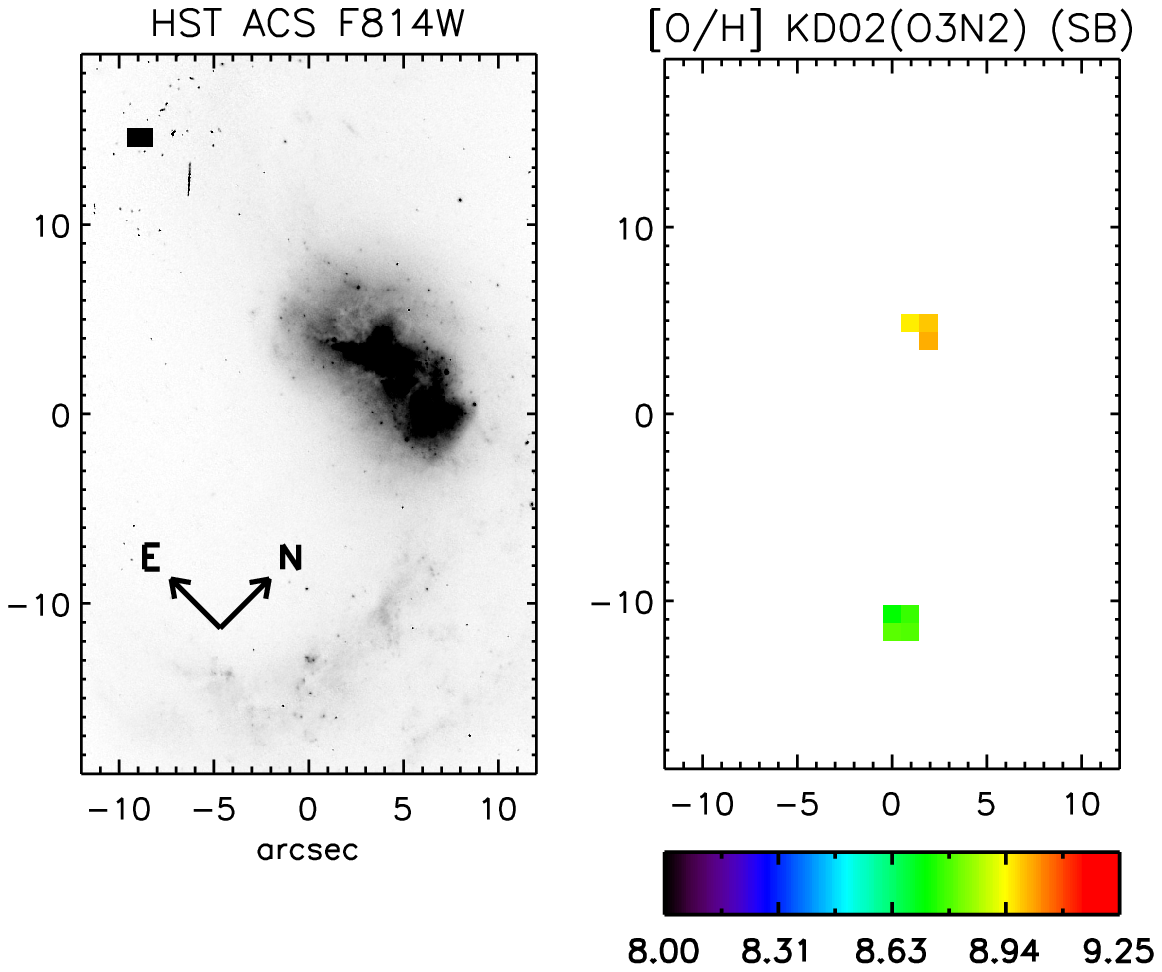}}
{\includegraphics[scale=0.99]{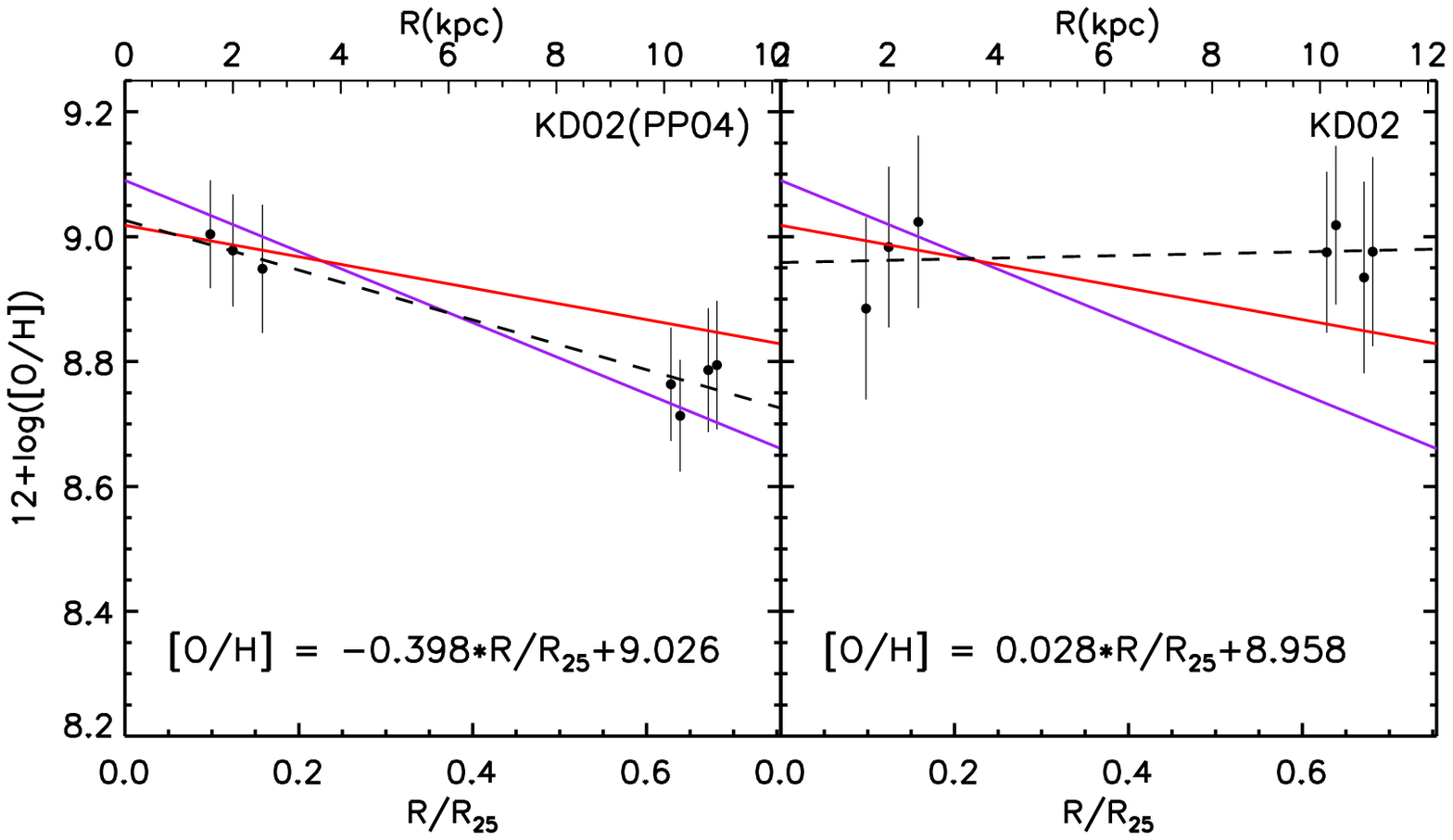}}
\caption{Same as fig A1 for IRAS F10038-3338}

\end{figure*}

\end{document}